\documentclass[aps,prb,reprint,superscriptaddress,longbibliography]{revtex4-2}
\usepackage{graphicx}
\usepackage{textcomp}
\usepackage{gensymb}
\usepackage{amsmath, amssymb}
\usepackage[colorlinks = true, citecolor = magenta]{hyperref}
\usepackage{braket}
\usepackage{natbib}
\usepackage[utf8]{inputenc}
\usepackage{float}
\begin{document}

\title{Role of inefficient measurement in realizing post-selection-based non-Hermitian qubits}

\author{Roson~Nongthombam}
\email{n.roson@iitg.ac.in}

\author{Aman~Verma}
\email{v.aman@iitg.ac.in}

\author{Amarendra~K.~Sarma}
\email{aksarma@iitg.ac.in}

\affiliation{Department of Physics, Indian, Institute of Technology Guwahati, Guwahati-781039, (India)}

\date{\today}

\begin{abstract}

Post-selecting against quantum jumps into the ground state $\ket{g}$ confines the evolution of the three-level system to the excited $\ket{e}$–$\ket{f}$ manifold, effectively realizing a $\mathcal{PT}$-symmetric non-Hermitian qubit. 
In this work, by introducing post-selection efficiencies for both decay channels, $\ket{f} \xrightarrow[]{} \ket{e}$ and $\ket{e} \xrightarrow[]{} \ket{g}$, we formulate a hybrid-Liouvillian framework that captures the unmonitored dynamics of this non-Hermitian qubit. 
We find that the decoherence effects arising from quantum jumps within the $\ket{e}$–$\ket{f}$ manifold also manifest under inefficient post-selection of $\ket{e}$–$\ket{f}$ transitions, thereby modifying the spectral properties of the Liouvillian and leading to a splitting of the exceptional points.
A comparative analysis shows that the trajectory-based approach—obtained by ensemble-averaging stochastic measurement trajectories generated via the Bayesian state update rule—and the ensemble-averaged Lindblad evolution remain consistent. 
% Moreover, due to the inherent non-Hermitian nature of the dynamics, the system exhibits a Zeno-like effect, reflected in an enhanced occupation of the excited state within the effective qubit manifold. 
Our results highlight the fundamental role of measurement inefficiency in realizing post-selection-based non-Hermitian qubits and in shaping the structure of Liouvillian exceptional points.  
These findings provide new insights into how inefficient measurement processes influence non-Hermitian behavior in open quantum systems.

\end{abstract}

\maketitle

%%%%%%%%%%%%%%%%%%%%%%%%%%%%%%%%%%%%%%%%%%%%%%%%%%%%%%%%%%
\section{Introduction}
%%%%%%%%%%%%%%%%%%%%%%%%%%%%%%%%%%%%%%%%%%%%%%%%%%%%%%%%%%

Non-Hermitian quantum systems with complex spectra may possess exceptional points (EPs), where eigenvalues coalesce and the corresponding eigenvectors become parallel \cite{Heiss_2012, Ashida02072020, doi:10.1142/S0219887810004816, PhysRevLett.80.5243}.
In recent years, the study of EPs in quantum theory has gained significant attention due to their applications in controlling quantum properties such as entanglement \cite{PhysRevA.100.063846, PhysRevLett.131.100202}, quantum state engineering \cite{PhysRevA.105.L010203}, and in various classical systems \cite{Peng2014, feng2017, hodaei2014, PhysRevLett.115.040402, xiao2017, peng}.
A particularly important framework where EPs arise naturally is that of parity-time ($\mathcal{PT}$) symmetric systems \cite{Naghiloo2019}.
Although $\mathcal{PT}$-symmetric Hamiltonians are generally non-Hermitian, they can exhibit entirely real spectra below a certain threshold, known as the $\mathcal{PT}$-symmetry unbroken phase.
This transition marks a non-Hermitian phase boundary with profound consequences for system dynamics and a range of applications \cite{hodaei2017, xu2016, zhang2017, PhysRevB.100.134505, shi2016, chen2017, lau2018}.
In quantum optics and open quantum systems, the emergence of Liouvillian exceptional points (LEPs)—considered to be the true quantum EPs \cite{PhysRevA.98.042118}—has been shown to drastically affect relaxation dynamics and coherence properties \cite{PhysRevResearch.5.043036, PhysRevLett.128.110402}.
% For an open system described by  Lindblad master equation with positive decay rates.  

In recent studies, the realization of a non-Hermitian system via post-selection of stochastic quantum measurements has been experimentally demonstrated in a three-level superconducting circuit QED platform \cite{Naghiloo2019, PhysRevLett.127.140504}.
Post-selection restricts the dynamics to the excited-state subspace by eliminating quantum jumps to the ground state, effectively realizing a non-Hermitian two-level system. 
In the full Lindblad description, spontaneous emission events transfer population out of the ${\ket{e}-\ket{f}}$ manifold into the ground state $\ket{g}$. 
By conditioning on the absence of such quantum jumps from $\ket{e}$ to $\ket{g}$ and $\ket{f}$ to $\ket{e}$, the system evolves under the no-jump Lindblad equation or the corresponding non-Hermitian effective Hamiltonian. 
In this regime, the eigenvalues coalesce at a third-order exceptional point (EP), marking the transition between oscillatory (PT-unbroken) and exponential (PT-broken) dynamics, consistent with the textbook non-Hermitian picture. 
On the other hand, when quantum jumps from $\ket{f}$ to $\ket{e}$ are included, the Liouvillian spectrum supports only a second-order EP. 
In this case, the transition from the unbroken to the broken $\mathcal{PT}$-symmetric phase manifests in the system's dynamics as a change from underdamped oscillatory behavior to an overdamped regime. 

In practice, it is difficult to perfectly post-select the quantum jump process. The hybrid-Liouvillian formalism, which explores the effects of inefficient post-selection, has been studied for a generic two-level system \cite{PhysRevA.101.062112}. 
In this work, we study the effects of inefficient measurement and the post-selection of quantum jumps from $\ket{e}$ to $\ket{g}$ and from $\ket{f}$ to $\ket{e}$.
We analyze the system dynamics using both the monitored stochastic quantum trajectory approach \cite{weber2014, lewalle2020,jordan2016,PhysRevA.96.053807, PhysRevA.88.042110, PhysRevA.92.032125, PhysRevX.6.011002, naghiloo2016, PhysRevA.96.022104, ficheux2018} and the unmonitored Liouvillian superoperator approach. 
In the trajectory approach, the Bayesian state update rule is applied to generate the evolution of the system state, and the ensemble average of many such trajectories is taken. 
The averaged dynamics obtained from this procedure are then compared with the dynamics obtained from the unmonitored Liouvillian formalism.
We discuss the inefficiency of post-selection in two scenarios: one where both the $\ket{f} \to \ket{e}$ and $\ket{e} \to \ket{g}$ jumps are post-selected (no-jump case), and the other where only the $\ket{e} \to \ket{g}$ jump is post-selected (jump case). 
We observe that inefficient post-selection in the no-jump case produces a Liouvillian eigenvalue spectrum and corresponding state evolution similar to that of the jump case. 
The comparison of both cases is analyzed using both monitored quantum trajectories and the unmonitored Liouvillian approach, with both approaches yielding consistent results. 
These findings provide new insights into how inefficient measurement processes influence non-Hermitian behavior in open quantum systems.

This paper is organized as follows: In Section~\ref{Sec:Model and Methods}, we describe the measurement scheme used to unravel the trajectories of the monitored dynamics of the three-level system. 
Next, we introduce the post-selection procedure and present the stochastic master equation describing the inefficient measurement process. 
We then show how a non-Hermitian qubit can be realized using post-selection. The trajectory and Liouvillian approaches are discussed for studying the unmonitored dynamics of the non-Hermitian qubit. 
In Section~\ref{Sec: Comparison}, we present a comparison between both approaches. Finally, we summarize our results and conclusions in Section~\ref{Sec: Results}.

%%%%%%%%%%%%%%%%%%%%%%%%%%%%%%%%%%%%%%%%%%%%%%%%%%%%%%%%%%
\section{Model and Methods}
\label{Sec:Model and Methods}
%%%%%%%%%%%%%%%%%%%%%%%%%%%%%%%%%%%%%%%%%%%%%%%%%%%%%%%%%%

%%%%%%%%%%%%%%%%%%%%%%%%%%%%%%%%%%%%%%%%%%%%%%%%%%%%%%%%%%
\subsection{Detecting the Spontaneous Emission of a Three-Level System}
\label{Sec:Detecting the Spontaneous Emission of a Three-Level System}
%%%%%%%%%%%%%%%%%%%%%%%%%%%%%%%%%%%%%%%%%%%%%%%%%%%%%%%%%%

\begin{figure}[t]
    \centering
    \includegraphics[width=0.4\textwidth]{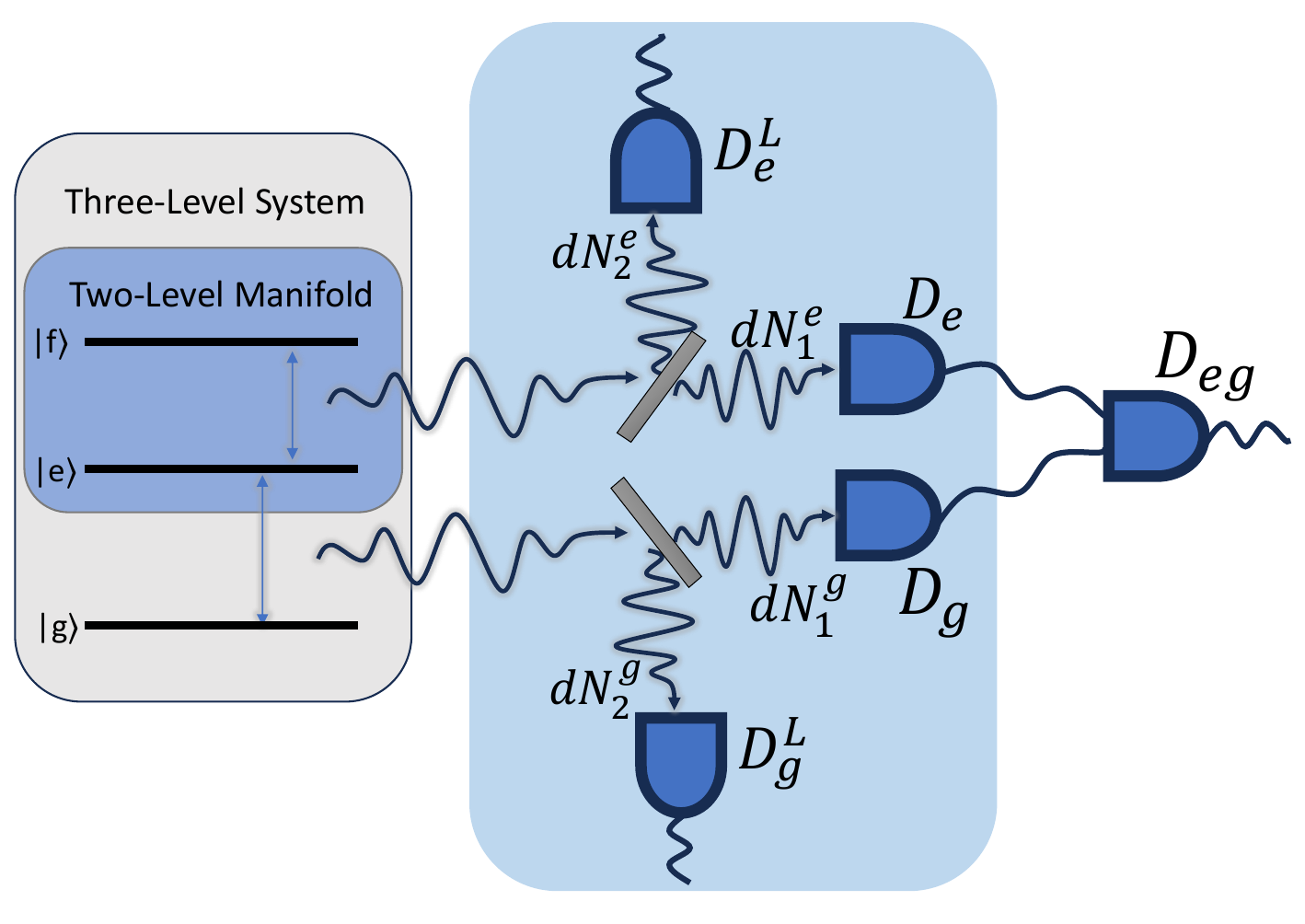}
    \caption{Schematic of the three-level system with monitored decay channels. 
    The excited states $\ket{f}$ and $\ket{e}$ decay via $\ket{f} \rightarrow \ket{e}$ 
    and $\ket{e} \rightarrow \ket{g}$ transitions with rates $\Gamma_e$ and $\Gamma_g$, 
    respectively. These decays are monitored by photodetectors $\mathcal{D}_e$ and 
    $\mathcal{D}_g$. The stochastic processes $dN_1^e$ and $dN_1^g$ register quantum 
    jumps at these detectors with detection efficiencies $\eta_e$ and $\eta_g$. 
    The processes $dN_e^L$ and $dN_g^L$ correspond to undetected jumps due to 
    imperfect efficiency ($1-\eta_{e,g}$). The direct transition $\ket{f} \rightarrow \ket{g}$ 
    is forbidden. The state of the system is updated based on the measurement outcome of detector $\mathcal{D}_{eg}$}
    \label{fig:Three_level_system}
\end{figure}

To study the dynamics of a non-Hermitian qubit, we consider a three-level system with states $\ket{f}$, $\ket{e}$, and $\ket{g}$, corresponding to the second excited, first excited, and ground states, respectively. 
The system undergoes spontaneous emission through the transitions $\ket{f} \xrightarrow{} \ket{e}$ and $\ket{e} \xrightarrow{} \ket{g}$, and the photons emitted from these decays are photodetected by a detector $D_{eg}$, as shown in Fig.~\ref{fig:Three_level_system}. 
The detector consists of two components that separately detect the photons emitted from the system. 
One component, $\mathcal{D}_e$, detects photons from the $\ket{f} \xrightarrow{} \ket{e}$ decay with efficiency $\eta_e$, while the other component, $\mathcal{D}_g$, monitors the $\ket{e} \xrightarrow{} \ket{g}$ decay with efficiency $\eta_g$. 
Since both components can be inefficient, the undetected photons are registered by corresponding loss detectors $\mathcal{D}^L_e$ and $\mathcal{D}^L_g$, with efficiencies $1-\eta_e$ and $1-\eta_g$, respectively.
Initially the state of the three-level system and the environment (detectors) are uncorrelated and can be written as 
\begin{align}
    \ket{\Psi(0)} = (p_g\,\ket{g} + p_e\,\ket{e} + p_f\,\ket{f}) \otimes \ket{D},
\end{align}
where $\ket{D}$ denotes the state of the environment or detector. 
The coefficients $|p_g|^2$, $|p_e|^2$, and $|p_f|^2$ represent the probabilities of finding the system in their respective states, according to Born’s rule. 
We further assume that the direct transition $\ket{f} \to \ket{g}$ is forbidden, i.e., such a transition does not occur. 
When spontaneous emission takes place, the environment captures the photon emitted from the system, which is then sent to the corresponding detector components. 
The phenomenological state update, according to Bayesian rule \cite{lewalle2020}, at time $t + dt$ is given by \cite{arXiv_2510_13345}
\begin{align}
    \ket{\Psi(t + dt)} =\; & p_e(t) \sqrt{\Gamma_g\, dt} \ket{g} \ket{D_{1g}} 
    + p_f(t) \sqrt{\Gamma_e\, dt} \ket{e} \ket{D_{1e}} \nonumber \\
    & + \Big( p_g(t) \ket{g} 
    + p_e(t) \sqrt{1-\Gamma_g\, dt} \ket{e} \nonumber \\
    & + p_f(t) \sqrt{1-\Gamma_e\, dt} \ket{f} \Big) \ket{D_0},
    \label{eq:state_update_equation}
\end{align}
Where $\ket{D_0}$, $\ket{D_{1e}}$, and $\ket{D_{1g}}$ denote the detector states. Specifically, $\ket{D_0}$ corresponds to the absence of a quantum jump, $\ket{D_{1e}}$ corresponds to a decay from $\ket{f}$ to $\ket{e}$ with probability $\Gamma_e~dt$, and $\ket{D_{1g}}$ corresponds to a decay from $\ket{e}$ to $\ket{g}$ with probability $\Gamma_g~dt$.
The parameters $\Gamma_e$ and $\Gamma_g$ represent the decay rates associated with the transitions $\ket{f} \to \ket{e}$ and $\ket{e} \to \ket{g}$, respectively.
We assume the detectors operate in the Markovian regime, such that $\Gamma_e\, dt \ll 1$ and $\Gamma_g\, dt \ll 1$. 
At time $t + dt$, if no photon is detected, the environment is projected onto the vacuum state $\ket{D_0}$, leading to the conditional state of the system $\ket{\Psi(t+dt)}_s = \braket{D_0|\Psi(t+dt)} = K_0 \ket{\Psi(t)}_s$, where $K_0$ denotes the Kraus operator corresponding to the no-jump process. Conversely, if a photon is emitted through the decay channel $\ket{f} \to \ket{e}$, the environment collapses onto $\ket{D_{1e}}$, and the system state updates as $\ket{\Psi(t+dt)}_s = \braket{D_{1e}|\Psi(t+dt)} = K_{1e}\ket{\Psi(t)}_s$, with $K_{1e}$ being the Kraus operator for this jump. Similarly, detection of a photon from the transition $\ket{e} \to \ket{g}$ projects the environment onto $\ket{D_{1g}}$, resulting in $\ket{\Psi(t+dt)}_s = \braket{D_{1g}|\Psi(t+dt)} = K_{1g}\ket{\Psi(t)}_s$, where $K_{1g}$ represents the Kraus operator for the $\ket{e} \to \ket{g}$ decay channel.
The Kraus operators $\{K_0,K_{1e},K_{1g}\}$ can be derived from a general krauss matrix
\begin{equation}
    M = \begin{bmatrix}
    \sqrt{1-\Gamma_edt} & 0 & 0 \\
    \sqrt{\Gamma_edt~} \hat{a}^\dagger_e & \sqrt{1-\Gamma_gdt} & 0 \\
    0 & \sqrt{\Gamma_gdt~} \hat{a}^\dagger_g & 1
    \end{bmatrix},
\end{equation}
where, $K_0 = \braket{D_0 | M | D_0}$, $K_{1g} = \braket{D_{1g} | M | D_0}$, and $K_{1e} = \braket{D_{1e} | M | D_0}$. The creation operators act on the environment state as $\hat{a}^\dagger_e \ket{D_0} = \ket{D_{1e}}$ and $\hat{a}^\dagger_g \ket{D_0} = \ket{D_{1g}}$ in the quantum trajectory picture. 
% Similar model has been explored with homodyne measurements scheme in \cite{arXiv:2510.13345}.   

\begin{figure}[ht]
    \centering
    \includegraphics[width=0.85\linewidth]{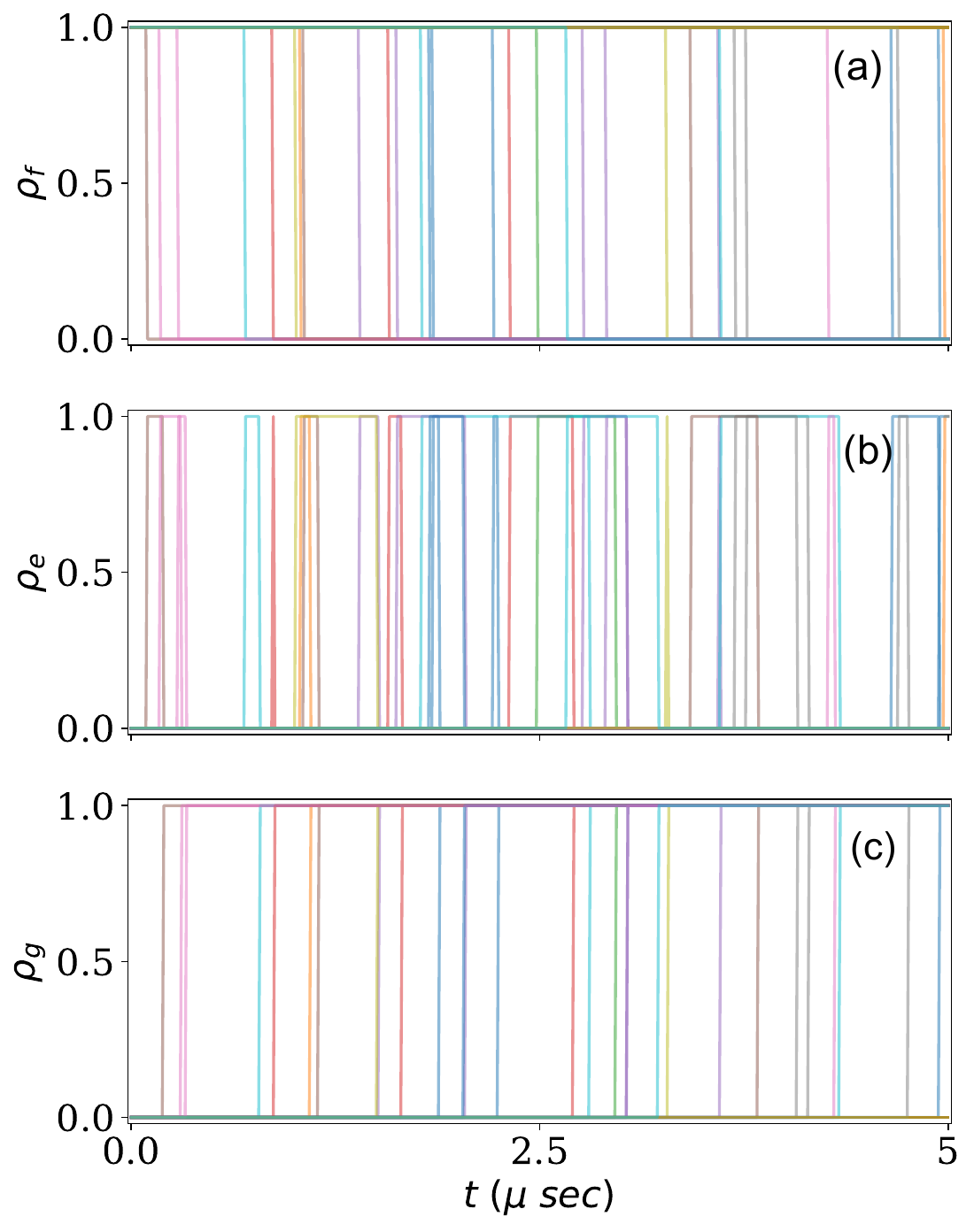}
    \includegraphics[width= 0.85 \linewidth]{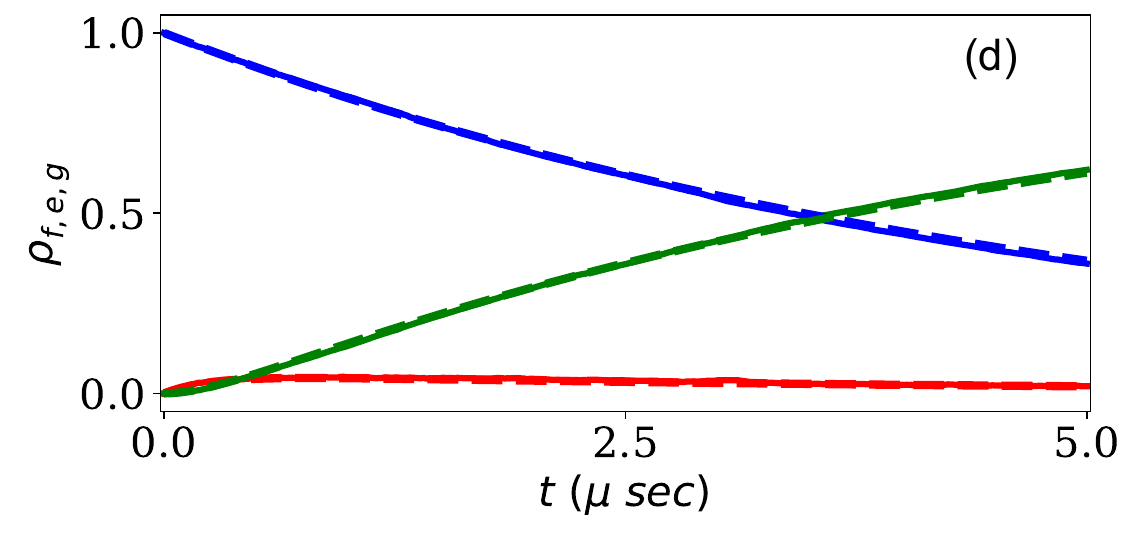}
    \caption{ Quantum trajectories and their ensemble average of a three-level system undergoing spontaneous emission. (a)–(c) Quantum jump trajectories of the system represented in terms of the populations of the states $\rho_{g,e,f}(t)$ as it undergoes spontaneous emission, or quantum jumps, from the excited states to the ground state, $\ket{f} \to \ket{e} \to \ket{g}$. There are 50 trajectories in the plot. Each trajectory evolves as follows: the system initially occupies the state $\ket{f}$, such that $\rho_f(t) = 1$ and $\rho_e(t) = \rho_g(t) = 0$, until a quantum jump from $\ket{f}$ to $\ket{e}$ occurs. After the jump, the system remains in the $\ket{e}$ state ($\rho_e(t) = 1$, $\rho_f(t) = \rho_g(t) = 0$) until a second quantum jump from $\ket{e}$ to $\ket{g}$ takes place. Once the system reaches the ground state $\ket{g}$, it stays there for the remainder of the evolution ($\rho_g(t) = 1$, $\rho_f(t) = \rho_e(t) = 0$).
    (d) The ensemble average of all the trajectories. The solid lines represent the average of $10^4$ trajectories, which matches perfectly with the Lindblad dynamics represented by the dashed lines. The blue, red, and green lines correspond to the average populations of the $\ket{f}$, $\ket{e}$, and $\ket{g}$ states, respectively.
    Parameters are $\Gamma_g = 4~\text{MHz}$ and $\Gamma_e = 0.2~\text{MHz}$.}
    \label{fig:Traj_plot}
\end{figure}

For inefficient detector components, the corresponding Kraus matrix can be written as
\begin{equation}
    M_{P} =
    \begin{bmatrix}
    \sqrt{1 - \Gamma_e dt} & 0 & 0 \\[5pt] 
    \begin{array}{c}
    \sqrt{(1 - \eta_e) \Gamma_e dt}~\hat{a}_{eL}^\dagger ~+ \\
    \sqrt{\eta_e \Gamma_e dt}\,\hat{a}_e^\dagger
    \end{array} & \sqrt{1 - \Gamma_g dt} & 0 \\
    0 & \begin{array}{c}
    \sqrt{\eta_g \Gamma_g dt}\,\hat{a}_g^\dagger ~+ \\[2pt]
    \sqrt{(1 - \eta_g) \Gamma_g dt}\,\hat{a}_{gL}^\dagger
    \end{array} & 1
    \end{bmatrix}
\label{eq:post_selection_Kraus}
\end{equation}
Here, the terms with $(1-\eta_e)$ and $(1-\eta_g)$ account for photons that are not detected by the detector components $\mathcal{D}_e$ and $\mathcal{D}_g$ due to inefficiency. 
These loss photons are instead captured by the corresponding loss detectors $\mathcal{D}_e^L$ and $\mathcal{D}_g^L$, with efficiencies $(1-\eta_e)$ and $(1-\eta_g)$, respectively. 
The corresponding Kraus operators are $\{K_P^{00,00}, K_P^{00,01}, K_P^{01,00}, K_P^{10,00}, K_P^{00,10}\}$, where the superscript $(\bullet\bullet, \bullet\bullet)$ denotes the clicks in the detectors $(\mathcal{D}_e~\mathcal{D}_e^L, \mathcal{D}_g~\mathcal{D}_g^L)$. In this notation, $0$ and $1$ correspond to the no-click and click events in the detectors, respectively. The state update of the system under inefficient detector components is as follows. When no photon is detected from either $\mathcal{D}_e$ or $\mathcal{D}_g$,
\begin{equation}
\begin{aligned}
    \rho(t + dt) = \frac{1}{N} \Big(& K_P^{\dagger\,(00,00)} \rho(t) K_P^{(00,00)} \\
    &+ K_P^{\dagger\,(01,00)} \rho(t) K_P^{(01,00)} \\
    &+ K_P^{\dagger\,(00,01)} \rho(t) K_P^{(00,01)} \Big),
    \label{eq:K00}
\end{aligned}
\end{equation}
with the normalization constant
\begin{equation}
\begin{aligned}
    N = \mathrm{Tr} \Big(& K_P^{\dagger\,(00,00)} \rho(t) K_P^{(00,00)} \\
    &+ K_P^{\dagger\,(01,00)} \rho(t) K_P^{(01,00)} \\
    &+ K_P^{\dagger\,(00,01)} \rho(t) K_P^{(00,01)} \Big).
\end{aligned}
\end{equation}
When the detectors $\mathcal{D}_e$ and $\mathcal{D}_g$ registers a click, the state update equations are given by
\begin{equation}
    \rho(t + dt) = \frac{K_P^{\dagger\,(10,00)} \rho(t) K_P^{(10,00)}}{\mathrm{Tr}\left(K_P^{\dagger\,(10,00)} \rho(t) K_P^{(10,00)}\right)},
    \label{eq:K10}
\end{equation}
and
\begin{equation}
    \rho(t + dt) = \frac{K_P^{\dagger\,(00,10)} \rho(t) K_P^{(00,10)}}{\mathrm{Tr}\left(K_P^{\dagger\,(00,10)} \rho(t) K_P^{(00,10)}\right)},
    \label{eq:K01}
\end{equation}
respectively.
The explicit expressions for the Kraus operators are provided in Appendix~\ref{appendix A}. 
Note that the state updates are governed by detections at the detector $D_{eg}$. 
The denominators in the state update equations represent the probabilities of the corresponding detection events under which the states are updated. 
By normalizing the probabilities of the three possible outcomes associated with the respective state updates, we evolve the system up to a total measurement time $T$. 
This procedure yields a single measurement trajectory of the system, as illustrated in Fig.~\ref{fig:Traj_plot}(a-c). 
We further compute the ensemble average over many such trajectories and plot the population dynamics of each energy level. 
For comparison, we also simulate the evolution using the Lindblad master equation for a three-level system undergoing spontaneous emission,
\begin{equation}
\frac{d\rho}{dt} = \Gamma_g\, \mathcal{D}(\ket{g}\bra{e})[\rho] + \Gamma_e\, \mathcal{D}(\ket{e}\bra{f})[\rho],
\label{eq:full_master_eq_three_level}
\end{equation}
where $\mathcal{D}(L)[\rho] = L\rho L^\dagger - \tfrac{1}{2}\{L^\dagger L, \rho\}$ denotes the standard Lindblad dissipator. The evolution obtained from the trajectory-based simulation agrees excellently with the results from the master equation approach as shown in Fig.~\ref{fig:Traj_plot}(d).

\begin{figure}[t]
    \centering
    \includegraphics[width=0.95\linewidth]{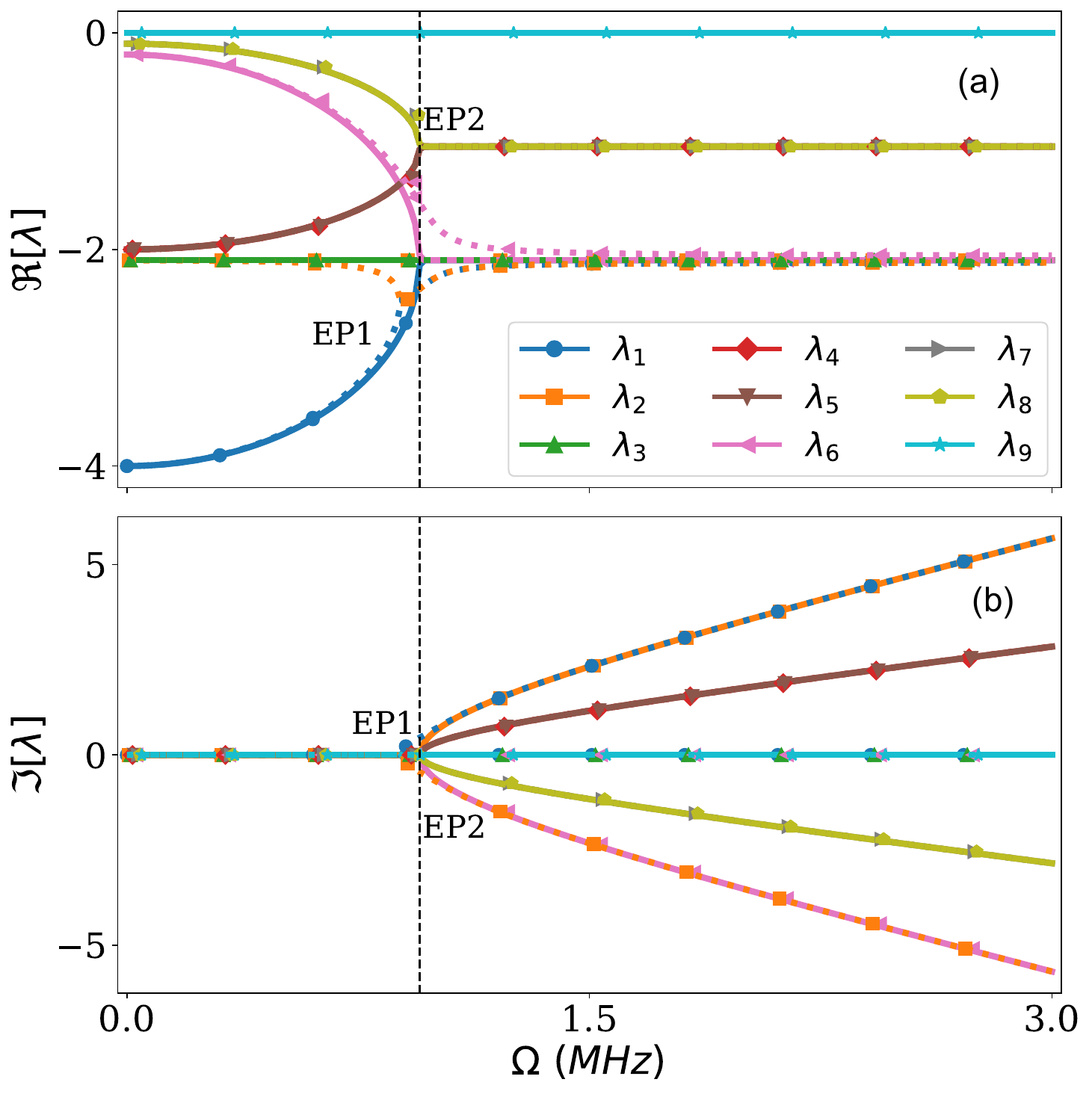}
    \includegraphics[width=0.95\linewidth]{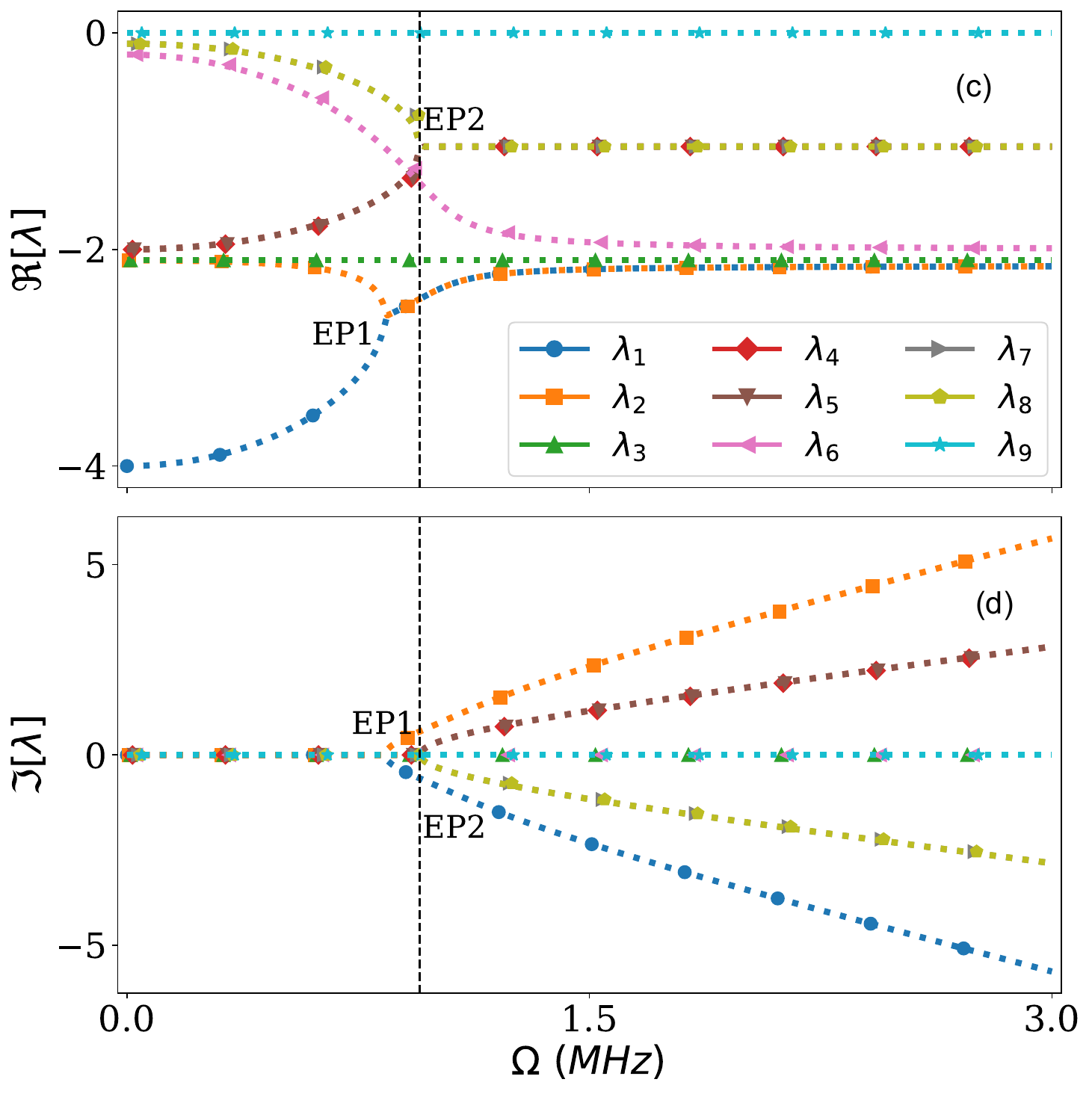}

    \caption{Real and imaginary parts of the Liouvillian eigenvalue spectrum as a function of $\Omega~\text{(MHz)}$ are shown for the no-jump (a)–(b) and jump (c)–(d) cases. In the no-jump case, solid lines correspond to $\eta_e = 1$, while dotted lines correspond to $\eta_e = 0.6$. For the jump case, the spectrum is independent of $\eta_e$. In both cases, $\eta_g = 1$. Other common parameters are $\Gamma_g = 4~\text{MHz}$ and $\Gamma_e = 0.2~\text{MHz}$.}
    \label{fig:liouvillian-pdf}
\end{figure}

%%%%%%%_________________________________________________________
\subsection{Non-Hermitian Qubit and the Post-selected Stochastic Master Equation}
%%%%%%%_________________________________________________________

The stochastic master equation corresponding to the inefficient measurement scheme discussed in section \ref{Sec:Detecting the Spontaneous Emission of a Three-Level System} is given as \cite{jacobs2006, Wiseman_Milburn_2009} 
% \begin{widetext}

\begin{align}
d\rho &= \big(-\tfrac{i}{\hbar}[H,\rho]
        - \tfrac{\Gamma_g}{2}\eta_g[\ket{e}\bra{e},\rho]_+
        + \eta_g \Gamma_g \braket{\ket{e}\bra{e}} \rho \notag \\ 
      &\quad + (1-\eta_g)\Gamma_g \,\mathcal{D}(\ket{g}\bra{e})[\rho]  
        - \tfrac{\Gamma_e}{2}\eta_e[\ket{f}\bra{f},\rho]_+ \notag \\ 
      &\quad  
        + \eta_e \Gamma_e \braket{\ket{f}\bra{f}} \rho  + (1-\eta_e)\Gamma_e \,\mathcal{D}(\ket{e}\bra{f})[\rho]\,\,\big) \,dt 
          \notag \\ 
      &\quad  + \left( \frac{\ket{e}\bra{f}\rho\ket{f}\bra{e}}
        {\braket{\ket{f}\bra{f}}} - \rho \right) dN_1^e \notag \\
       &\quad + \left( \frac{\ket{g}\bra{e}\rho\ket{e}\bra{g}}
        {\braket{\ket{e}\bra{e}}} - \rho \right) dN_1^g.
    \label{eq:full_master_eq}
\end{align}

where $[\bullet, \rho]_+ = \bullet \rho + \rho \bullet$ denotes the anticommutation bracket, and 
$\mathcal{D}(\hat{A})[\bullet] = \hat{A} \bullet \hat{A}^\dagger - \frac{1}{2}[\hat{A}^\dagger \hat{A}, \bullet]_+$ is the Lindblad dissipator. 
Here, $dN_1^g$ and $dN_1^e$ represent the counting processes for detector components $D_g$ and $D_e$, respectively, which take the value 1 if a photon is detected in the interval $dt$, and 0 otherwise. 
Similarly, $dN_2^g$ and $dN_2^e$ correspond to the loss detectors $D_g^L$ and $D_e^L$. 
The mean values of these counting processes are
\begin{alignat}{1}
\braket{dN_1^g} &= \Gamma_g\, \eta_g \braket{\ket{e}\bra{e}}\,dt, \\
\braket{dN_2^g} &= \Gamma_g\, (1-\eta_g) \braket{\ket{e}\bra{e}}\,dt, \label{dN_2^g} \\
\braket{dN_1^e} &= \Gamma_e\, \eta_e \braket{\ket{f}\bra{f}}\,dt, \label{eq:jump_f_e}\\
\braket{dN_2^e} &= \Gamma_e\, (1-\eta_e) \braket{\ket{f}\bra{f}}\,dt. \label{dN_2^e}
\end{alignat}
Since we assume the photon detected at the loss-detctors are lost, the corresponding signals are averaged out in Eq.~\eqref{eq:full_master_eq}.
% These are adifferentials of the Poisson process can either be $0$ or $1$, and the Ito rule holds $dN_i(t)\,dN_j(t)=\delta_{ij}\,dN_i(t)$.
Suppose we consider only the photocount signals measured from the detector $D_e$ and take the ensemble average of these measurement signals. The resulting average dynamics corresponds to that of a non-Hermitian qubit in the $\ket{f}$–$\ket{e}$ manifold. In other words, if we postselect all trajectories that reach the ground state or undergo a quantum jump to the ground state, the ensemble-averaged dynamics resembles that of a non-Hermitian qubit. This can be seen from Eq.~\eqref{eq:full_master_eq} by setting $dN_1^{g} = 0$ and averaging over $dN_1^e$. The resulting master equation then describes a non-Hermitian qubit with $\ket{f}$ as the excited state and $\ket{e}$ as the ground state.

A similar dynamics of the non-Hermitian qubit can also be obtained using the state update equations, Eqs.~\eqref{eq:K00}, \eqref{eq:K10}, and \eqref{eq:K01}. By taking the ensemble average over trajectories that do not undergo a quantum jump to the ground state, one obtains dynamics consistent with the master equation approach.

In the following sections, we derive the Liouvillian of the postselected non-Hermitian qubit and compare its dynamics with the ensemble-averaged trajectories. We consider two cases: one in which the spontaneous emission from $\ket{f}$ to $\ket{e}$ is post-selected (no-jump), and another where this jump is included (jump). We also vary the efficiency parameters of the detector components to study how the dynamics changes under different detection conditions.

%%%%%%%_________________________________________________________
\subsection{Liouvillian of the No-Jump Case}
\label{Sec:No-Jump Case}
%%%%%%%_________________________________________________________

The stochastic master equation for the post-selected non-Hermitian qubit can be obtained from Eq.~\eqref{eq:full_master_eq} by setting $dN_1^g = 0$ and omitting the normalization term $\eta_g \Gamma_g \braket{\ket{e}\bra{e}} \rho\, dt$. It takes the form
\begin{equation}
\begin{aligned}
d\rho &= -\tfrac{i}{\hbar}[H,\rho]\,dt
        - \tfrac{\Gamma_g}{2}\eta_g[\ket{e}\bra{e},\rho]_+\,dt \\
      &\quad  + (1-\eta_g)\Gamma_g \,\mathcal{D}(\ket{g}\bra{e})[\rho] \,dt - \tfrac{\Gamma_e}{2}\eta_e[\ket{f}\bra{f},\rho]_+\,dt \\
      &\quad  
        + \eta_e \Gamma_e \braket{\ket{f}\bra{f}} \rho \,dt  + (1-\eta_e)\Gamma_e \,\mathcal{D}(\ket{e}\bra{f})[\rho] \,dt\\ 
      &\quad 
        + \left( \frac{\ket{e}\bra{f}\rho\ket{f}\bra{e}}
        {\braket{\ket{f}\bra{f}}} - \rho \right) dN_1^e .
        \label{eq:postselect_SMQ}
\end{aligned}
\end{equation}
In the no-jump case, where the spontaneous emission from $\ket{f}$ to $\ket{e}$ is post-selected, we set $dN_1^e = 0$ and remove the normalization term $\eta_e \Gamma_e \braket{\ket{f}\bra{f}} \rho\, dt$. The Liouvillian corresponding to this no-jump evolution is then given by
% Now we consider the no-jump case for both the photodetectors which implies $dN_1^g=dN_1^e=0$ on taking the ensemble average of Eq.~\ref{eq:full_master_eq} and using Eq.~(\ref{eq:post_selection_Kraus}) up to the first order in $dt$ \cite{PhysRevA.88.042110,PhysRevA.96.053807} we get the Liouville operator as
\begin{align}
\mathcal{L}_{NJ}[\rho] ={}& -i[H, \rho]  
- \tfrac{\Gamma_g}{2}\eta_g \,[\ket{e}\bra{e}, \rho]_+  
   - \tfrac{\Gamma_e}{2}\eta_e \,[\ket{f}\bra{f}, \rho]_+ \nonumber \\
&+ (1-\eta_g)\Gamma_g \Big(\ket{g}\bra{e}\,\rho \,\ket{e}\bra{g} 
     - \tfrac{1}{2}[\ket{e}\bra{e}, \rho]_+\Big) \nonumber \\
&+ (1-\eta_e)\Gamma_e \Big(\ket{e}\bra{f}\,\rho \,\ket{f}\bra{e} 
     - \tfrac{1}{2}[\ket{f}\bra{f}, \rho]_+\Big).
\end{align} 
Note that for the case of perfect detection, that is, $\eta_g=\eta_e=1$, the non-Hermitian qubit evolves by the effective Hamiltonian 
\begin{align}
    H_{eff} = H - i\,\frac{\Gamma_g}{2} \ket{e}\bra{e} -i\,\frac{\Gamma_e}{2} \ket{f}\bra{f}.
    \label{eq:eff_hamiltonian}
\end{align}
Here, $H = \Omega\, \left(\ket{e}\bra{f} + \ket{f}\bra{e}\right)$ with $\Omega$ being the drive strength of the $\ket{f}$ and $\ket{e}$ states.
When the efficiencies are zero, we obtain the Lindblad master equation of the three-level system given by Eq.\eqref{eq:full_master_eq_three_level}.

In Fig.~(\ref{fig:liouvillian-pdf}), we plot the eigenvalue spectrum of the Liouvillian superoperators given in Appendix~\ref{Appendix: Liouville Superoperator}.
The solid lines correspond to the case of perfect post-selection, where $\eta_g = \eta_e = 1$ in $\mathcal{L}_{\mathrm{NJ}}$ in no-jump Liouville superoperator $\mathcal{L}_{NJ}$ Eq.~(\ref{eq:no_jump_liouvillian}). 
In this case, two exceptional points (EPs) are observed: a third-order exceptional point (EP$_1$) and a second-order exceptional point (EP$_2$), as indicated in the figure.
We also plot the eigenvalue spectrum for the case where the post-selection of the $\ket{f} \rightarrow \ket{e}$ transition is imperfect while that of $\ket{e} \rightarrow \ket{g}$ remains perfect, i.e., for $\eta_e <1$ and $\eta_g = 1$. 
This case is represented by the dotted curves in the figure. Here, both EP$_1$ and EP$_2$ are of second order. 
% Note that when $\eta_e <1$ in $\mathcal{L_{NJ}} $, $EP_1$ will become second order EP.  
Comparing the two cases, we find that the third-order EP$_1$ observed under perfect post-selection splits into a second-order EP when the detection efficiency is reduced. Later, we will show that the transition from a third-order to a second-order exceptional point, caused by imperfect post-selection, leads to decoherence within the $\ket{f}$ and $\ket{e}$ manifolds.
% Similar results were discussed in \cite{PhysRevA.100.062131}.

%%%%%%%_________________________________________________________
\subsection{Liouvillian of the Jump Case}
\label{Sec:Jump Case}
%%%%%%%_________________________________________________________

The Liouvillian of the post-selected non-Hermitian qubit, where the spontaneous emission from $\ket{f}$ to $\ket{e}$ is not post-selected, can be obtained from Eq.~\eqref{eq:postselect_SMQ} by averaging over the detector $D_e$ signal. 
The resulting master equation is
% To get the effect of quantum jump from $\ket{f}$ to $\ket{e}$ state, i.e, with only $dN_1^g = 0$ and from Eq.~\ref{eq:full_master_eq}, the master equation as 
\begin{align}
\mathcal{L}_J [\rho] ={}& -i[H, \rho]  - \tfrac{\Gamma_g}{2}\eta_g \,[\ket{e}\bra{e}, \rho]_+  \nonumber \\
   % - \tfrac{\Gamma_e}{2} \,[\ket{f}\bra{f}, \rho]_+ 
&+ (1-\eta_g)\Gamma_g \Big(\ket{g}\bra{e}\,\rho \,\ket{e}\bra{g} 
     - \tfrac{1}{2}[\ket{e}\bra{e}, \rho]_+\Big) \nonumber \\
&+ \Gamma_e \Big(\ket{e}\bra{f}\,\rho \,\ket{f}\bra{e} 
     - \tfrac{1}{2}[\ket{f}\bra{f}, \rho]_+\Big),
\label{eq:jump}
\end{align} 
where, we have substituted $\braket{dN_1^e} = \Gamma_e\, \eta_e\braket{\ket{f}\bra{f}}\,dt$.  
Notice that the Liouville operator given in Eq.~\eqref{eq:jump} doesn't depend on the efficiency $\eta_e $.
A perfect post-selection of the $\ket{e} \to \ket{g}$ quantum jumps can be realized by setting $\eta_g = 1$. When the efficiency is zero, the Lindblad master equation of the three-level system given by Eq.\eqref{eq:full_master_eq_three_level}.
% In this regime, the dynamics are effectively constrained to 
% the upper subspace, allowing one to isolate the role of the competing 
% decay channel $\ket{f} \to \ket{e}$. 
The superoperator form of the Liouville operators is given in Appendix~(\ref{Appendix: Liouville Superoperator}). 
We plot the eigenvalue spectrum of the Liouvillian superoperators in Fig.~\ref{fig:liouvillian-pdf} for the case of perfect post-selection, $\eta_g = 1$. We observe two second-order exceptional points, denoted by EP1 and EP2. 
The eigenvalues corresponding to EP2 are responsible for the decoherence within the $\ket{f}$–$\ket{e}$ manifold. A similar Liouvillian spectrum has been studied in Ref.~\cite{PhysRevLett.127.140504}.
Comparing the jump Liouvillian with the imperfect no-jump Liouvillian, we find that both exhibit similar spectral features characterized by second-order exceptional points. However, the splitting of the exceptional points from third to second order is more pronounced in the jump case than in the no-jump case. 
This difference results in distinct decoherence rates for the two scenarios~\cite{PhysRevLett.127.140504}, with the jump case exhibiting a higher rate of decoherence. 
Further discussion on this is provided in the next section.

%%%%%%%%%%%%%%%%%%%%%%%%%%%%%%%%%%%%%%%%%%%%%%%%%%%%%%%%
\section{Comparison between Trajectory approach and the Liouville Dynamics}
\label{Sec: Comparison}
%%%%%%%%%%%%%%%%%%%%%%%%%%%%%%%%%%%%%%%%%%%%%%%%%%%%%%%%

\begin{figure}[ht]
    \includegraphics[width=0.225\textwidth]{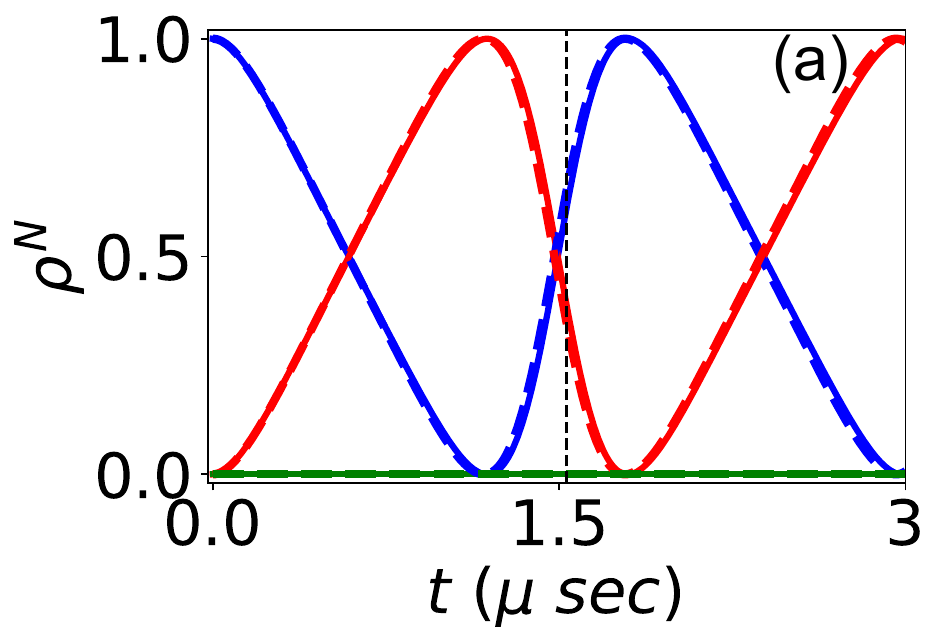}
    \includegraphics[width=0.225\textwidth]{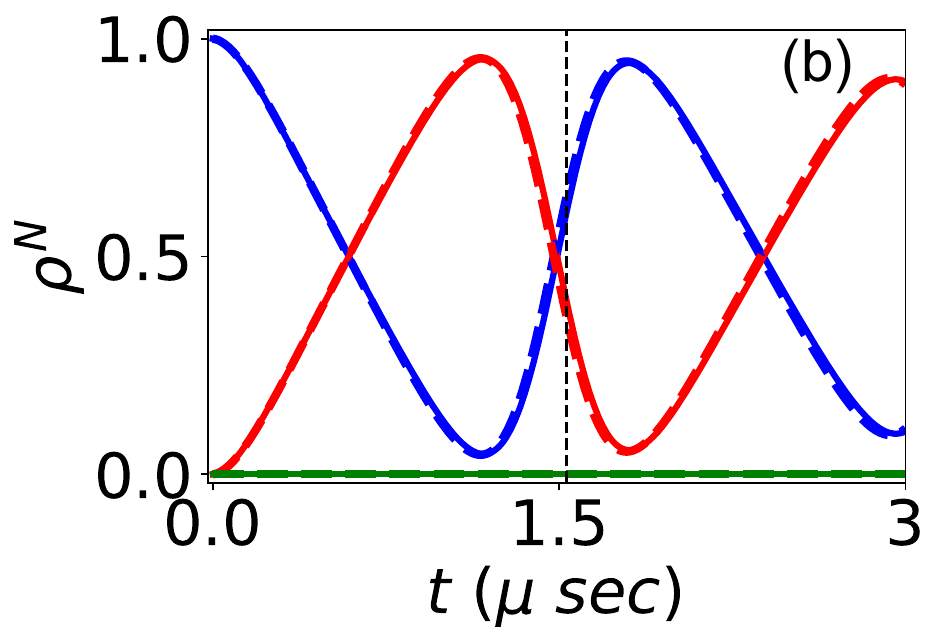}
    \includegraphics[width=0.225\textwidth]{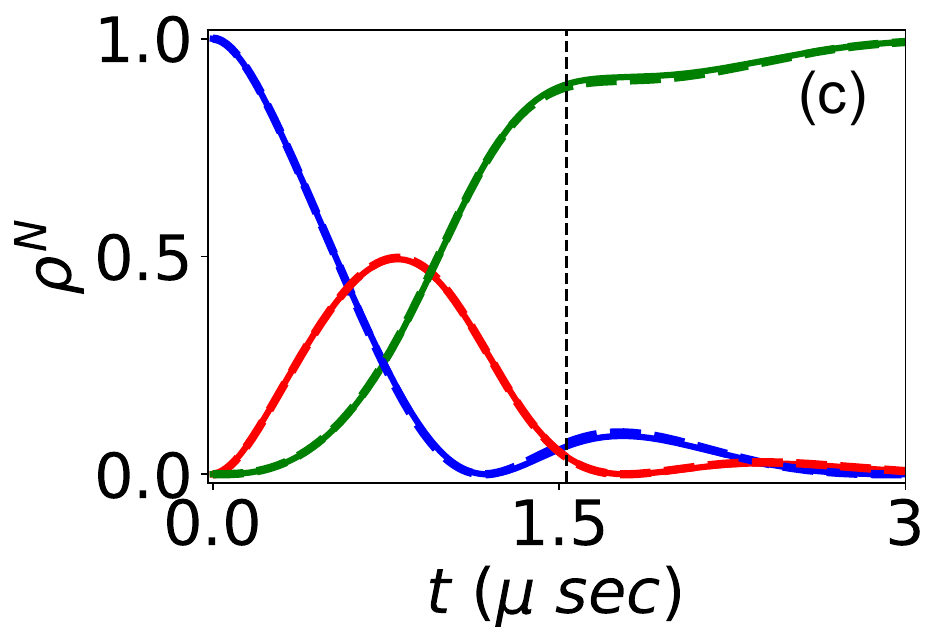}
    \includegraphics[width=0.225\textwidth]{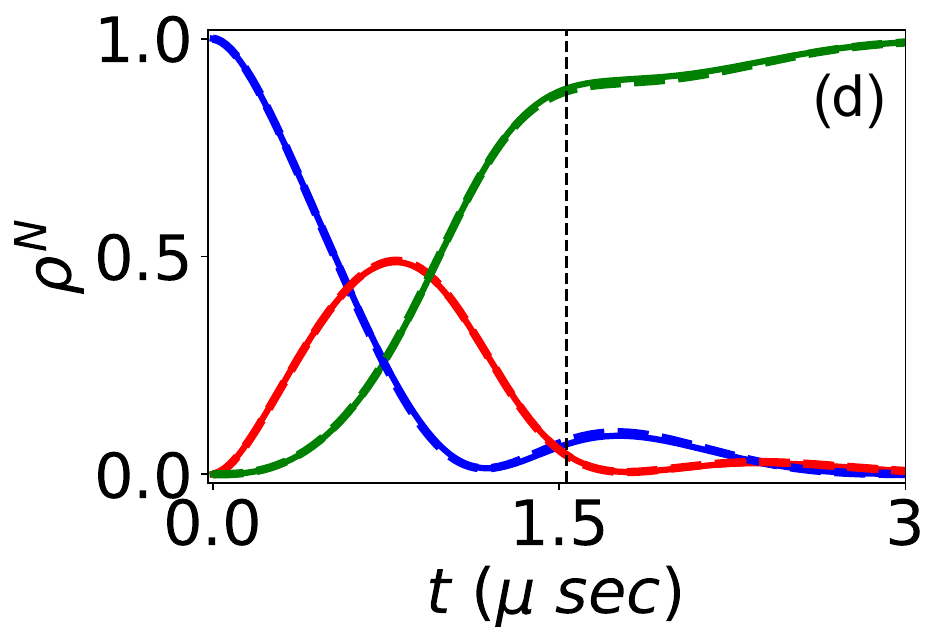}
    \caption{Evolution of the state populations for the no-jump case where post-selection is applied to both the $\ket{f} \rightarrow \ket{e}$ and $\ket{e} \rightarrow \ket{g}$ jumps. Dashed lines show the normalized populations derived from Liouvillian dynamics. Solid lines represent the population derived from average trajectory dynamics, with blue, red, and green corresponding to the populations of $\ket{f}$, $\ket{e}$, and $\ket{g}$ states, respectively. 
    (a) corresponds to the post-selection efficiencies $\eta_g = \eta_e = 1$. For inefficient detection, $\eta_e = 0.75$ with $\eta_g = 1$ is shown in (b). Panels (c) and (d) correspond to $\eta_g = 0.75, \eta_e = 1$ and $\eta_g = 0.75, \eta_e = 0.75$, respectively. The black dotted line indicates the time period of coherent oscillation of the population dynamics, $t_c = \pi / \mathrm{Re}[\Phi_0] = 1.61~\mu\mathrm{s}$, where $\Phi_0$ is the first eigenvalue of the non-Hermitian Hamiltonian in Eq.~(\ref{eq:eff_hamiltonian}). 
    Simulations were performed with $10^5$ trajectories. Other parameters are 
    $\Gamma_g = 4~\mathrm{MHz}$ and $\Gamma_e = 0.2~\mathrm{MHz}$, $\Omega = 2~\mathrm{MHz}$ with.
    }

    \label{fig:No_jump_2}
\end{figure}

\begin{figure}[ht]
    \includegraphics[width=0.235\textwidth]{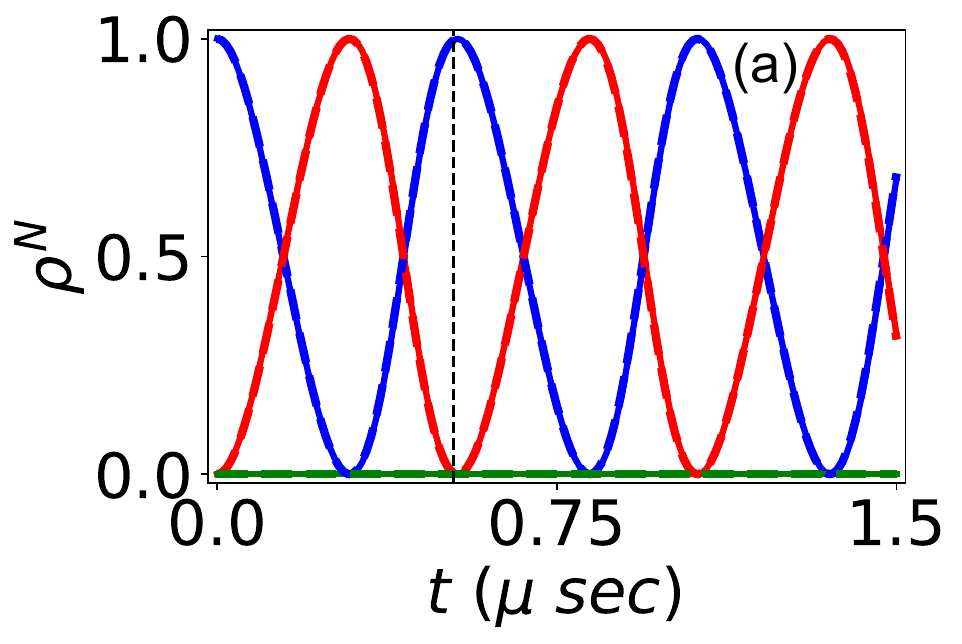}
    \includegraphics[width=0.235\textwidth]{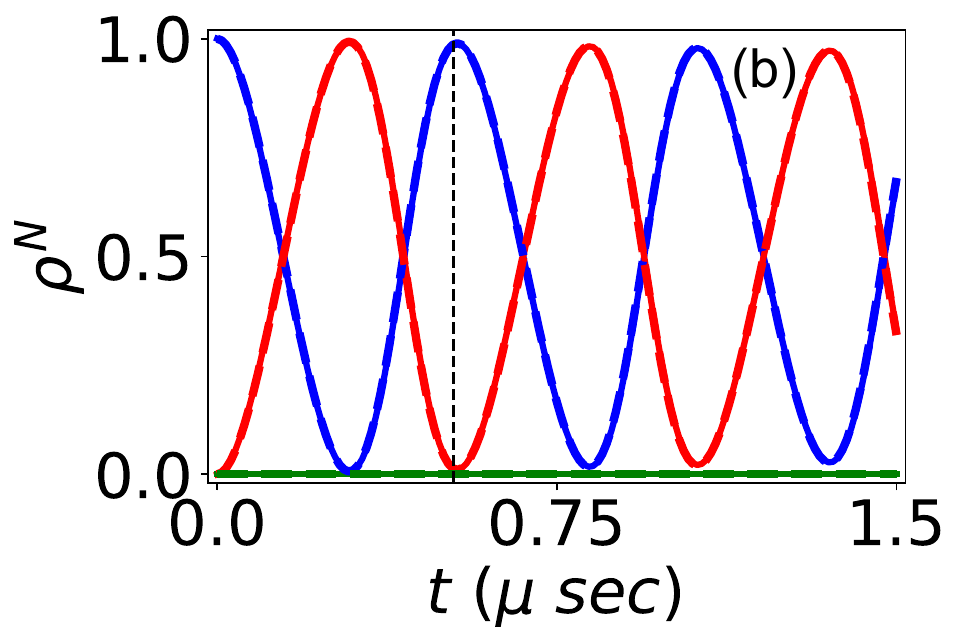}
    \includegraphics[width=0.235\textwidth]{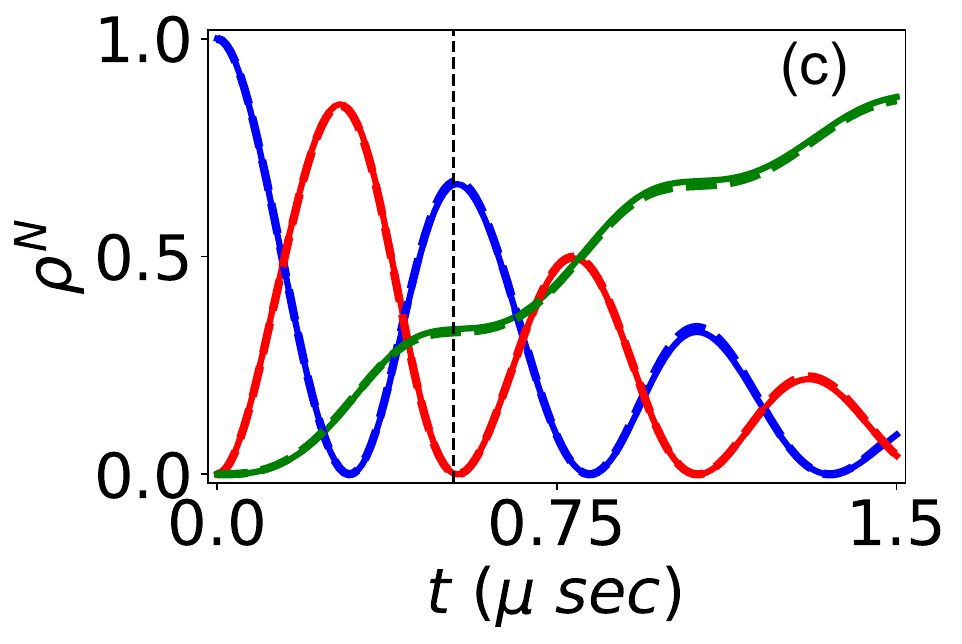}
    \includegraphics[width=0.235\textwidth]{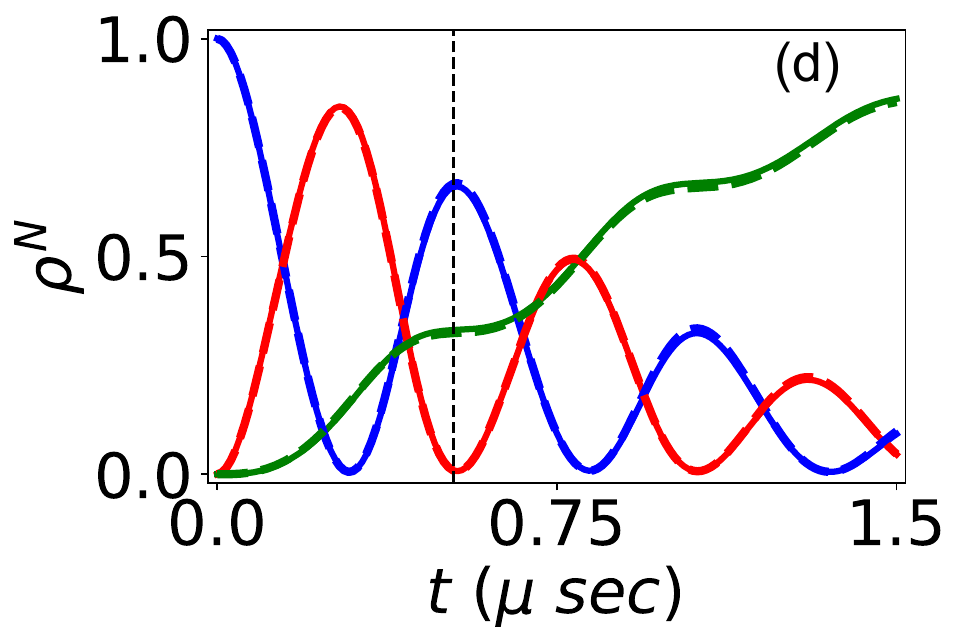}

    \label{fig:No_jump_J_6}
    \caption{
    Evolution of the state populations for the no-jump case for $\Omega = 6~\mathrm{MHz}$. Dashed lines show the normalized populations derived from Liouvillian dynamics. Solid lines represent the population derived from average trajectory dynamics, with blue, red, and green corresponding to the populations of $\ket{f}$, $\ket{e}$, and $\ket{g}$ states, respectively. 
    (a) corresponds to the post-selection efficiencies $\eta_g = \eta_e = 1$. For inefficient detection, $\eta_e = 0.75$ with $\eta_g = 1$ is shown in (b). Panels (c) and (d) correspond to $\eta_g = 0.75, \eta_e = 1$ and $\eta_g = 0.75, \eta_e = 0.75$, respectively. The black dotted line indicates the time period of coherent oscillation of the population dynamics, $t_c = 0.52\mu\mathrm{s}$.
    Simulations were performed with $10^5$ trajectories. Other parameters are 
    $\Gamma_g = 4~\mathrm{MHz}$ and $\Gamma_e = 0.2~\mathrm{MHz}$.
    } 
\end{figure}

We compare the evolution of the state populations obtained from the ensemble average of the trajectories generated using the state update relations discussed in Section~\ref{Sec:Detecting the Spontaneous Emission of a Three-Level System} with those obtained from the Liouvillian dynamics described in Sections~\ref{Sec:Jump Case} and~\ref{Sec:No-Jump Case}. 

First, we consider the case where both the $\ket{f}\!\to\!\ket{e}$ and $\ket{e}\!\to\!\ket{g}$ quantum jumps are postselected (the \textit{no-jump case}). 
In Fig.~\ref{fig:No_jump_2}, we present the evolution of the state populations derived from the 
trajectory average and the normalized populations,
\begin{equation}
\rho_i^N = \frac{\rho_i}{\rho_g + \rho_e + \rho_f}, \quad (i = g, e, f),
\end{equation}
obtained from the Liouvillian dynamics for four different scenarios at a drive strength close to EP2.
(1) Perfect post-selection of both $\ket{f}\!\to\!\ket{e}$ and $\ket{e}\!\to\!\ket{g}$ transitions (Fig.~\ref{fig:No_jump_2}(a)). 
In this case, the ground-state population remains zero, and due to the unitary drive within the $\ket{f}$–$\ket{e}$ manifold, we observe coherent oscillations between these two states. 
Note that, because of the post-selection, the system tends to remain predominantly in the $\ket{f}$ state. 
(2) Imperfect post-selection of the $\ket{f}\!\to\!\ket{e}$ transition and perfect post-selection of the $\ket{e}\!\to\!\ket{g}$ transition  (Fig.~\ref{fig:No_jump_2}(b)). 
Here, the imperfect post-selection leads to decoherence in the oscillations. 
The decoherence rate corresponds to the splitting of the exceptional point (EP) from third to second order, as shown in Fig.~\ref{fig:liouvillian-pdf}(a). 
(3) Perfect post-selection of the $\ket{f}\!\to\!\ket{e}$ transition but imperfect post-selection of the $\ket{e}\!\to\!\ket{g}$ transition  (Fig.~\ref{fig:No_jump_2}(c)). 
In this case, coherent oscillations between $\ket{f}$ and $\ket{e}$ are observed. 
However, due to the imperfect post-selection in the $\ket{e}\!\to\!\ket{g}$ channel, the ground-state population gradually increases, leading to decay into the ground state at longer times. 
(4) Imperfect post-selection of both $\ket{f}\!\to\!\ket{e}$ and $\ket{e}\!\to\!\ket{g}$ transitions  (Fig.~\ref{fig:No_jump_2}(d)). 
The behavior is qualitatively similar to that in case (c), with the additional effect that the oscillations between $\ket{f}$ and $\ket{e}$ become less coherent due to the inefficiency in $\eta_e$.

In the no-jump case, we also plot the four scenarios for a drive strength away from the exceptional point EP2. We observe behavior similar to that seen near EP2. 
In this case, the oscillations are more pronounced, and the $\ket{f}$ and $\ket{e}$ states have equal probabilities, unlike in the near-EP2 case where $\ket{f}$ is preferentially occupied.
In both plots, the black dotted line in all population graphs represents the time period 
\(
t_c = \frac{\pi}{\mathrm{Re}[\Phi_0]},
\)
where $\Phi_0$ is the first eigenvalue of the non-Hermitian Hamiltonian in Eq.~(\ref{eq:eff_hamiltonian}). This time period corresponds to the coherent oscillation of the population dynamics \cite{Naghiloo2019}.

\begin{figure}[t]
    \centering
    \includegraphics[width=0.235\textwidth]{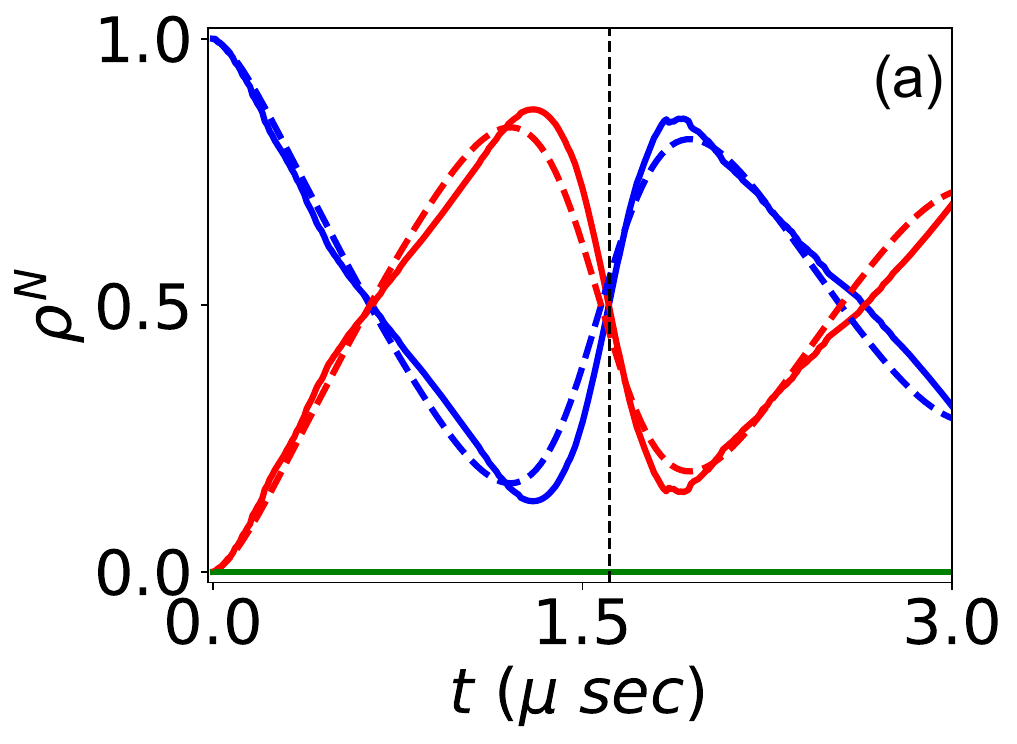}
    \includegraphics[width=0.235\textwidth]{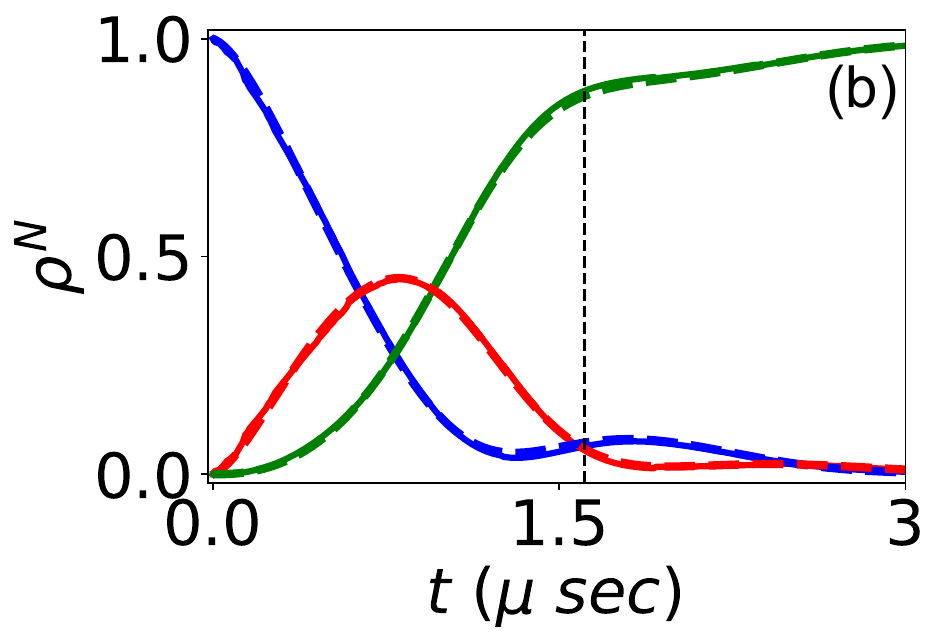}
    \caption{Evolution of the state populations for the jump case where post-selection is applied only to $\ket{e} \rightarrow \ket{g}$ jumps. Dashed lines show the normalized populations derived from Liouvillian dynamics. Solid lines represent the population derived from average trajectory dynamics, with blue, red, and green corresponding to the populations of $\ket{f}$, $\ket{e}$, and $\ket{g}$ states, respectively. 
    (a) corresponds to the post-selection efficiencies $\eta_g = 1$. For inefficient detection or post-selection $\eta_g = 0.75$ is shown in (b). Here $t_c = 1.61~\mu\mathrm{s}$. Other parameters are 
    $\Gamma_g = 4~\mathrm{MHz}$ and $\Gamma_e = 0.2~\mathrm{MHz}$, $\Omega = 2~\mathrm{MHz}$.}
    \label{fig:Jump J= 2}
\end{figure}

\begin{figure}[t]
    \centering
    \includegraphics[width=0.235\textwidth]{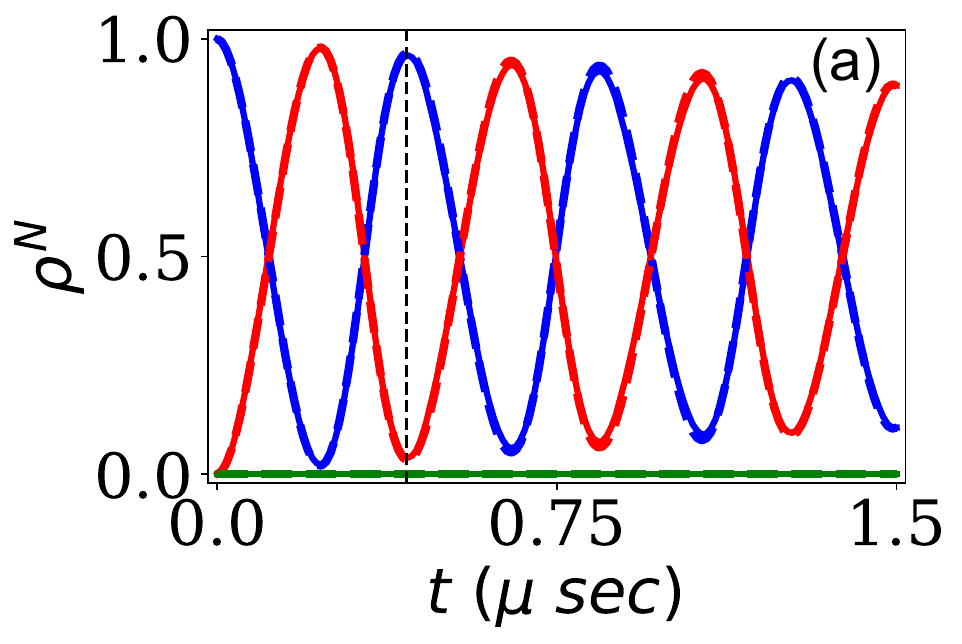}
    \includegraphics[width=0.235\textwidth]{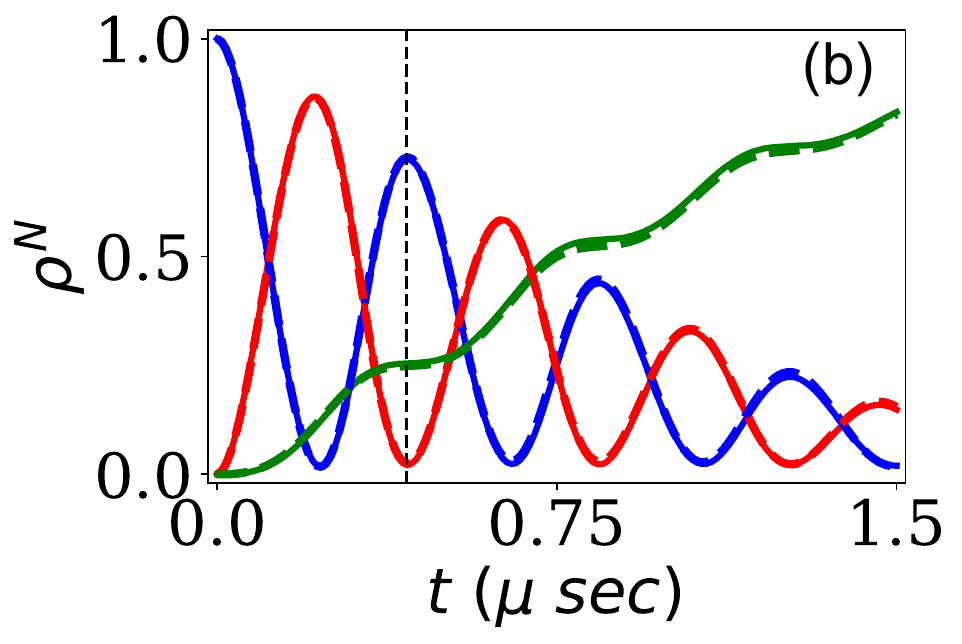}
    \caption{Evolution of the state populations for the jump case for $\Omega = 7.5~\mathrm{MHz}$. Dashed lines show the normalized populations derived from Liouvillian dynamics. Solid lines represent the population derived from average trajectory dynamics, with blue, red, and green corresponding to the populations of $\ket{f}$, $\ket{e}$, and $\ket{g}$ states, respectively. 
    (a) corresponds to the post-selection efficiencies $\eta_g = 1$. (b) For inefficient detection $\eta_g = 0.75$. The black dotted line indicates the time period of coherent oscillation of the population dynamics, $t_c = 0.42\mu\mathrm{s}$. Other parameters are $\Gamma_g = 4~\mathrm{MHz}$ and $\Gamma_e = 0.2~\mathrm{MHz}$.
     }
    \label{fig:Jump J= 7.5}
\end{figure}

Now, we consider the \textit{jump case}, where the $\ket{f}\!\to\!\ket{e}$ transition is not postselected, while the $\ket{e}\!\to\!\ket{g}$ quantum jumps are postselected. We show the evolution of the normalized populations for two drive strengths: one close to EP2 and the other away from EP2. 
First, we consider the scenarios at a drive strength close to EP2: 
(1) Perfect post-selection of the $\ket{e}\!\to\!\ket{g}$ jump. 
Here, we observe behavior similar to scenario (2) of the no-jump case. However, in this case, the decoherence is caused not by imperfect post-selection of the $\ket{f}\!\to\!\ket{e}$ transition, but due to the direct quantum jump. 
As shown in Fig.~\ref{fig:liouvillian-pdf}(c), the splitting of eigenvalues at EP2 is larger than in the inefficiency case shown in Fig.~\ref{fig:liouvillian-pdf}(a), resulting in increased decoherence of the oscillations, as illustrated in Fig.~\ref{fig:Jump J= 2}.  
(2) Imperfect post-selection of the $\ket{e}\!\to\!\ket{g}$ jump. 
This scenario is similar to scenario (4) of the no-jump case. Here, the decoherence is stronger due to the effect of the direct quantum jump.
At a drive strength away from EP2, we observe behavior similar to that of the near-EP2 case. 
However, in this regime, the oscillations are more pronounced, as shown in Fig.~\ref{fig:Jump J= 7.5}.

\begin{figure}[t]
    \centering
    \includegraphics[width=0.235\textwidth]{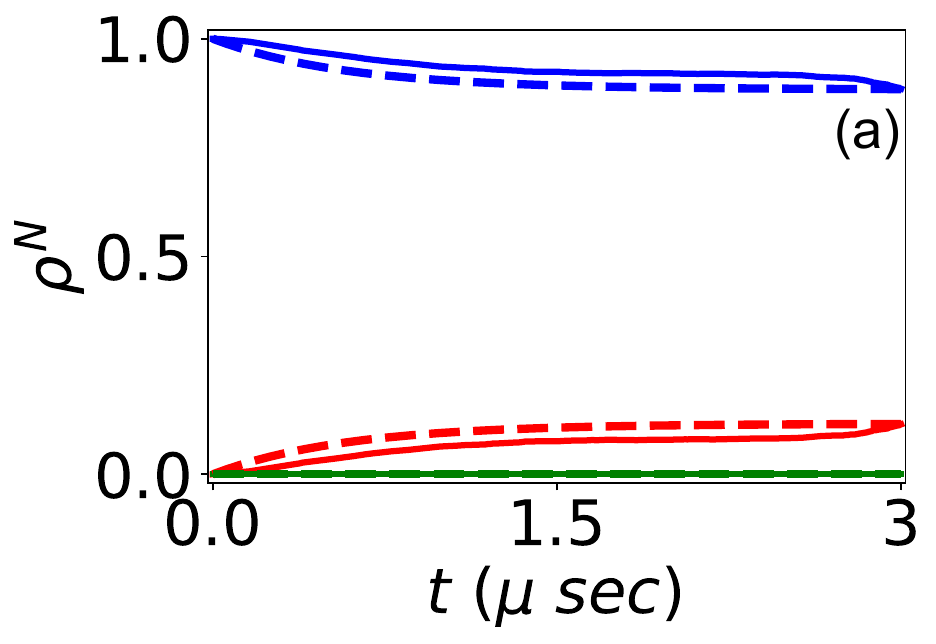}
    \includegraphics[width=0.235\textwidth]{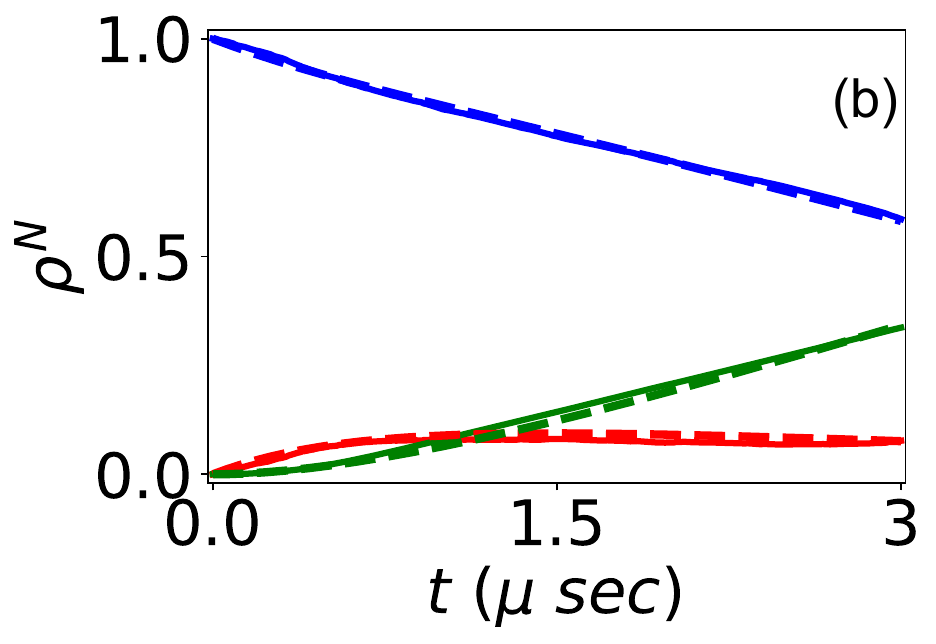}
    \caption{Evolution of the state populations for the jump case for $\Omega = 0.5~\mathrm{MHz}$. Dashed lines show the normalized populations derived from Liouvillian dynamics. Solid lines represent the population derived from average trajectory dynamics, with blue, red, and green corresponding to the populations of $\ket{f}$, $\ket{e}$, and $\ket{g}$ states, respectively. 
    (a) corresponds to the post-selection efficiencies $\eta_g = 1$. (b) For inefficient detection $\eta_g = 0.75$. Other parameters are $\Gamma_g = 4~\mathrm{MHz}$ and $\Gamma_e = 0.2~\mathrm{MHz}$.
     }
    \label{fig:Jump J= 0.5}
\end{figure}

At a drive strength below the exceptional point, the non-Hermitian qubit undergoes overdamped oscillations, since all eigenvalues of the Liouvillian are real and negative. The normalized populations at this drive strength for the jump case, with perfect and imperfect post-selection of the $\ket{e}\!\to\!\ket{g}$ jump, are shown in Fig.~\ref{fig:Jump J= 0.5}. In the perfect post-selection case, the ground-state population remains zero, and the evolution within the $\ket{f}$–$\ket{e}$ manifold is overdamped. In the imperfect post-selection case, the $\ket{f}$–$\ket{e}$ manifold still exhibits overdamped oscillations, but the population of the ground state increases due to the inefficient post-selection. Consequently, the system eventually decays to the ground state.

%----------------------------------------
\section{Conclusion}
\label{Sec: Results}
In conclusion, we have studied the effects of inefficient measurement and post-selection of quantum jumps in a three-level system, focusing on jumps from $\ket{e}$ to $\ket{g}$ and from $\ket{f}$ to $\ket{e}$. 
Using both monitored stochastic quantum trajectories and the unmonitored Liouvillian formalism, we show that ensemble-averaged trajectory dynamics are consistent with Liouvillian evolution. 
We observe that inefficient post-selection in the no-jump case produces a Liouvillian eigenvalue spectrum and corresponding state evolution similar to that of the jump case. 
Furthermore, decoherence effects arising from quantum jumps within the $\ket{e}$–$\ket{f}$ manifold also manifest under inefficient post-selection of $\ket{e}$–$\ket{f}$ transitions, thereby modifying the Liouvillian spectrum and leading to a splitting of the exceptional points. 
These results highlight the fundamental role of measurement backaction and detection inefficiency in shaping non-Hermitian dynamics, providing important insights into the influence of realistic measurement processes on the behavior of open quantum systems.

% We have presented a comprehensive comparative analysis between the dynamics of ensemble-averaged trajectories and the Lindblad dynamics for a three-level system subject to inefficient monitoring. By selectively excluding the spontaneous $\ket{e}\rightarrow\ket{g}$ transitions, the system dynamics become restricted to the $\ket{f}$–$\ket{e}$ manifold, effectively realizing a two-level non-Hermitian qubit. We have studied the effect of in effect 
% Upon including the $\ket{f}\rightarrow\ket{e}$ decay channel, we observe that the trajectory-based evolution and the full Lindblad description deviate significantly, especially in the vicinity of the Liouville exceptional point (LEP). This mismatch originates from the non-orthogonality of the Liouvillian eigenmodes near coalescence, where stochastic measurement backaction strongly influences the transient population dynamics. However, when the system operates at Rabi frequencies far from the LEP or when the detection efficiency satisfies $\eta_g < 1$, both approaches converge and exhibit nearly identical behavior, indicating the suppression of non-Hermitian effects and the dominance of conventional dissipative evolution.

% Our analysis emphasize the essential role of measurement backaction and detection inefficiency in controlling the interplay between coherent and dissipative processes. This framework establishes a continuous interpolation between purely non-Hermitian and fully Lindbladian dynamics, providing a unified perspective on how quantum measurements shape exceptional-point physics in open systems.

\section*{Acknowledgement}
This work is supported by MoE, Government of India (Grant No. MoE-STARS/STARS-2/2023-0161). A.V. gratefully acknowledges a research fellowship from MoE, Government of India.

\appendix
\section{Kraus operators for $M_P$}
\label{appendix A}

Here, we provide details of the Kraus operators associated with the post-selection scheme. The general form of the Kraus operator is written as \begin{equation}
K_P^{(\mathcal{D}_e~\mathcal{D}_e^L,\mathcal{D}_g~\mathcal{D}_g^L)} = 
\langle \mathcal{D}_e~\mathcal{D}_e^L,\mathcal{D}_g~\mathcal{D}_g^L | 
M_P | 
\mathcal{D}_{0e}~\mathcal{D}_{0e}^L,\mathcal{D}_{0g}~\mathcal{D}_{0g}^L \rangle,
\end{equation}
where $\hat{a}_e^\dagger|\mathcal{D}_{0e}\rangle = |\mathcal{D}_{1e}\rangle$, 
$\hat{a}_{eL}^\dagger|\mathcal{D}_{0e}^L\rangle = |\mathcal{D}_{1e}^L\rangle$, 
$\hat{a}_g^\dagger|\mathcal{D}_{0g}\rangle = |\mathcal{D}_{1g}\rangle$, and 
$\hat{a}_{gL}^\dagger|\mathcal{D}_{0g}^L\rangle = |\mathcal{D}_{1g}^L\rangle$.  
In this notation, the subscripts $0$ and $1$ correspond to no-click and click events in the detectors, respectively.
The explicit forms of the Kraus operators are given below.

\textbf{(1) No photons detected in either detector:}
The Kraus operator is obtained from $K_P^{(\mathcal{D}_{0e}~\mathcal{D}_{0e}^L,\mathcal{D}_{0g}~\mathcal{D}_{0g}^L)}=\langle\mathcal{D}_{0e}~\mathcal{D}_{0e}^L,\mathcal{D}_{0g}~\mathcal{D}_{0g}^L|M_P|\mathcal{D}_{0e}~\mathcal{D}_{0e}^L,\mathcal{D}_{0g}~\mathcal{D}_{0g}^L\rangle$ or 
\begin{equation}
K_{P}^{00,00} =
\begin{bmatrix}
\sqrt{1 - \Gamma_edt} & 0 & 0 \\ 

0 & \sqrt{1 - \Gamma_g} & 0 \\
0 & 0 & 1
\label{eq:00,00}
\end{bmatrix}.
\end{equation}
Here $K_P^{(\mathcal{D}_{0e}~\mathcal{D}_{0e}^L,\mathcal{D}_{0g}~\mathcal{D}_{0g}^L)}=K_P^{(00,00)}$.

\textbf{(2) $\ket{e}\!\to\!\ket{g}$ jump detected at $\mathcal{D}_g^L$:}
The Kraus operator is obtained from $K_P^{(\mathcal{D}_{0e}~\mathcal{D}_{0e}^L,\mathcal{D}_{0g}~\mathcal{D}_{1g}^L)}=\langle\mathcal{D}_{0e}~\mathcal{D}_{0e}^L,\mathcal{D}_{0g}~\mathcal{D}_{1g}^L|M_P|\mathcal{D}_{0e}~\mathcal{D}_{0e}^L,\mathcal{D}_{0g}~\mathcal{D}_{0g}^L\rangle$ or 

\begin{equation}
K_{P}^{00,01} =
\begin{bmatrix}
0 & 0 & 0 \\ 
0 & 0 & 0 \\
0 & \sqrt{\Gamma_g~dt(1-\eta_g)} & 0
\label{eq:00,01}
\end{bmatrix}.
\end{equation}
Here $K_P^{(\mathcal{D}_{0e}~\mathcal{D}_{0e}^L,\mathcal{D}_{0g}~\mathcal{D}_{1g}^L)}=K_P^{(00,01)}$.

\textbf{(3) $\ket{f}\!\to\!\ket{e}$ jump detected at $\mathcal{D}_e^L$:}
The Kraus operator is obtained from
$K_P^{(\mathcal{D}_{0e}~\mathcal{D}_{1e}^L,\mathcal{D}_{0g}~\mathcal{D}_{0g}^L)}=\langle\mathcal{D}_{0e}~\mathcal{D}_{1e}^L,\mathcal{D}_{0g}~\mathcal{D}_{0g}^L|M_P|\mathcal{D}_{0e}~\mathcal{D}_{0e}^L,\mathcal{D}_{0g}~\mathcal{D}_{0g}^L\rangle$ or
\begin{equation}
K_{P}^{01,00} =
\begin{bmatrix}
0 & 0 & 0 \\ 
\sqrt{\Gamma_e dt\,(1-\eta_e)} & 0 & 0 \\
0 & 0 & 0
\label{eq:01,00}
\end{bmatrix}.
\end{equation}
Here $K_P^{(\mathcal{D}_{0e}~\mathcal{D}_{0e}^L,\mathcal{D}_{0g}~\mathcal{D}_{1g}^L)}=K_P^{(01,00)}$.

\textbf{(4) $\ket{e}\!\to\!\ket{g}$ jump detected at $\mathcal{D}_g$:}
The Kraus operator is obtained from
$K_P^{(\mathcal{D}_{0e}~\mathcal{D}_{0e}^L,\mathcal{D}_{1g}~\mathcal{D}_{0g}^L)}=\langle\mathcal{D}_{0e}~\mathcal{D}_{0e}^L,\mathcal{D}_{1g}~\mathcal{D}_{0g}^L|M_P|\mathcal{D}_{0e}~\mathcal{D}_{0e}^L,\mathcal{D}_{0g}~\mathcal{D}_{0g}^L\rangle$ or
\begin{equation}
K_{P}^{10,00} =
\begin{bmatrix}
0 & 0 & 0 \\ 
\sqrt{\Gamma_e dt\,\eta_e} & 0 & 0 \\
0 & 0 & 0
\label{eq:10,00}
\end{bmatrix}.
\end{equation}
Here $K_P^{(\mathcal{D}_{0e}~\mathcal{D}_{0e}^L,\mathcal{D}_{0g}~\mathcal{D}_{1g}^L)}=K_P^{(00,10)}$.

\textbf{(5) $\ket{f}\!\to\!\ket{e}$ jump detected at $\mathcal{D}_e$:}
The Kraus operator is obtained from
$K_P^{(\mathcal{D}_{1e}~\mathcal{D}_{0e}^L,\mathcal{D}_{0g}~\mathcal{D}_{0g}^L)}=\langle\mathcal{D}_{1e}~\mathcal{D}_{0e}^L,\mathcal{D}_{0g}~\mathcal{D}_{0g}^L|M_P|\mathcal{D}_{0e}~\mathcal{D}_{0e}^L,\mathcal{D}_{0g}~\mathcal{D}_{0g}^L\rangle$ or
\begin{equation}
K_{P}^{00,10} =
\begin{bmatrix}
0 & 0 & 0 \\ 
0 & 0 & 0 \\
0 & \sqrt{\Gamma_g~dt\eta_g} & 0
\label{eq:00,10}
\end{bmatrix}.
\end{equation}
Here $K_P^{(\mathcal{D}_{1e}~\mathcal{D}_{0e}^L,\mathcal{D}_{0g}~\mathcal{D}_{1g}^L)}=K_P^{(10,00)}$.

%%%%%%%%%%%%%%%%%%%%%%%%%%%%%%%%%%%%%%%%%%%%%%%%%%%%%%%%%
\section{Spectral Properties of Liouville Super Operator}
\label{Sec:Spectral Properties of Liouville Super Operator}
%%%%%%%%%%%%%%%%%%%%%%%%%%%%%%%%%%%%%%%%%%%%%%%%%%%%%%%%%%%%

In the theory of open quantum systems, the Lindblad (GKLS) master equation is the most widely applicable master equation in the Markovian limit \cite{10.1093/acprof:oso/9780199213900.001.0001}: 
\begin{align}
    \partial_t \hat{\rho}(t) &= -i[\hat{H}, \hat{\rho}(t)] \nonumber \\
    &\quad + \sum_k \Gamma_k \left( \hat{L}_k \hat{\rho}(t) \hat{L}_k^\dagger 
    - \tfrac{1}{2}\{\hat{L}_k^\dagger \hat{L}_k, \hat{\rho}(t)\} \right)
    \label{eq:General_master_eq}
\end{align}
Eq.~(\ref{eq:General_master_eq}) can be written in terms of Liouvillian super operator 
\begin{align}
    \partial_t\hat{\rho}(t) = \mathcal{L}\hat{\rho}(t)
\end{align}

Now we define the left-handed $\hat{\phi}_i$ and right-handed $\hat{\psi}_i$ eigenvectors of the Liouville superoperator as 
\begin{align}
   \mathcal{L}\hat{\phi}_i = \lambda_i\hat{\phi}_i \\ 
   \mathcal{L}^\dagger\hat{\psi}_i = \lambda_i^*\hat{\psi}_i
\end{align}
where $\lambda_i's$ are the eigenvalues of the Liouville superoperator and $Tr[\hat{\phi}_k\hat{\psi}_l] =\delta_{k,l}$. Moreover, these eigenvalues $\mathcal{L}$ encode the essential features of open-system dynamics. The steady state is associated with the zero eigenvalue, $\lambda_0=0$, while all other eigenvalues satisfy $\mathrm{Re}(\lambda_k)\leq 0$, ensuring physical decay. For $\lambda_k=\alpha_k+i\beta_k$, the real part $\alpha_k$ sets the decay rate and the imaginary part $\beta_k$ gives the oscillation frequency (see Fig.~(\ref{fig:liouvillian-pdf})). Due to the orthonormality condition, the density matrix $\hat{\rho}(t)$ can be expended in terms of Liouville eigenvectors as \cite{Naghiloo2019}. 

\begin{align}
    \hat{\rho}(t) = \sum c_j(t)\,\hat{\phi}_j.
\end{align}

where $c_j(t) = exp(\lambda_jt)Tr\left[(\hat{\phi}_j)\hat{\rho}(0)\right]$. The dynamics generated by the above Liouville formalism are compared with those obtained from the quantum trajectory method in the main text. In Fig.~(\ref{fig:liouvillian-pdf}), we show the real and imaginary parts of the Liouville eigenvalues for the case of efficient and inefficient post-section. It is observe that, by decreasing the post-selection efficiencies, one can shift the location of the EP and even transform its nature, i.e., from a third-order EP to a second-order EP\cite{PhysRevLett.127.140504, Naghiloo2019}. This clearly demonstrates the crucial effects of quantum jump in exceptional points. Moreover, the real part of the Liouville eigenvalues are negative, $\alpha_k= \Re{[\lambda_k]}<0, k=1,2~...~9$, directly implying the positivity of the decay rates in the matrix equation and hence Markovian dynamics \cite{CHRUSCINSKI20221}.

\section{Liouville Super Operators}
\label{Appendix: Liouville Superoperator}

\subsection{Liouville Super Operator for No-Jump Case}

\begin{widetext}
\begin{equation}
\mathcal{L}_{NJ} =
\left(
\begin{array}{ccccccccc}
 -\Gamma_e & i \Omega  & 0 & -i \Omega  & 0 & 0 & 0 & 0 & 0 \\
 i \Omega  & -\tfrac{\Gamma_e}{2}-\tfrac{\Gamma_g}{2} & 0 & 0 & -i \Omega  & 0 & 0 & 0 & 0 \\
 0 & 0 & -\tfrac{\Gamma_e}{2} & 0 & 0 & -i \Omega  & 0 & 0 & 0 \\
 -i \Omega  & 0 & 0 & -\tfrac{\Gamma_e}{2}-\tfrac{\Gamma_g}{2} & i \Omega  & 0 & 0 & 0 & 0 \\
 \Gamma_e (1-\eta_e) & -i \Omega  & 0 & i \Omega  & -\Gamma_g & 0 & 0 & 0 & 0 \\
 0 & 0 & -i \Omega  & 0 & 0 & -\tfrac{\Gamma_g}{2} & 0 & 0 & 0 \\
 0 & 0 & 0 & 0 & 0 & 0 & -\tfrac{\Gamma_e}{2} & i \Omega & 0 \\
 0 & 0 & 0 & 0 & 0 & 0 & i \Omega  & -\tfrac{\Gamma_g}{2} & 0 \\
 0 & 0 & 0 & 0 & \Gamma_g (1-\eta_g) & 0 & 0 & 0 & 0 \\
\end{array}
\right)
\label{eq:no_jump_liouvillian}
\tag{A1}
\end{equation}
\end{widetext}

\subsection{Liouville Super operator for Jump Case}

\begin{widetext}
\begin{equation}
\mathcal{L}_{J} =
    \left(
            \begin{array}{ccccccccc}
             -\text{$\Gamma_e $} & i \Omega  & 0 & -i \Omega  & 0 & 0 & 0 & 0 & 0 \\
             i \Omega  & -\frac{\text{$\Gamma_e $}}{2}-\frac{\text{$\Gamma_g $}}{2} & 0 & 0 & -i \Omega  & 0 & 0 & 0 & 0 \\
             0 & 0 & -\frac{\text{$\Gamma_e$}}{2} & 0 & 0 & -i \Omega  & 0 & 0 & 0 \\
             -i \Omega  & 0 & 0 & -\frac{\text{$\Gamma_e $}}{2}-\frac{\text{$\Gamma_g$}}{2} & i \Omega  & 0 & 0 & 0 & 0 \\
             \text{$\Gamma_e$} & -i \Omega  & 0 & i \Omega  & -\text{$\Gamma_g$} & 0 & 0 & 0 & 0 \\
             0 & 0 & -i \Omega  & 0 & 0 & -\frac{\text{$\Gamma_g$}}{2} & 0 & 0 & 0 \\
             0 & 0 & 0 & 0 & 0 & 0 & -\frac{\text{$\Gamma_e$}}{2} & i \Omega  & 0 \\
             0 & 0 & 0 & 0 & 0 & 0 & i \Omega  & -\frac{\text{$\Gamma_g$}}{2} & 0 \\
        0 & 0 & 0 & 0 & \text{$\Gamma_g$} (1-\text{$\eta_g $}) &        0  & 0 & 0 & 0
        \end{array}
    \right)
\label{eq:Jump_Liouville_superoperator}
\end{equation}
\end{widetext}

%-----------------------------------------

\newpage
\bibliography{ref}

\end{document}